\documentclass[a4paper,11pt]{article}

\usepackage[top=1.2in,right=0.5in,left=0.5in,bottom=1.2in]{geometry}
 
\usepackage{graphics,graphicx}
\usepackage{float}
\usepackage{caption}
\usepackage{subcaption}
\usepackage{booktabs}
\usepackage{longtable}
\usepackage{supertabular}
\usepackage{tabularx}
\usepackage{multirow}
\usepackage{array}
\usepackage{tabu}
\usepackage{mathtools}

\usepackage{hyperref}

\usepackage[ruled,vlined]{algorithm2e}

\usepackage{stackengine,scalerel}
\usepackage{accents}

\def\salto#1#2{
\left[\mbox{\hspace{-#1em}}\left[#2\right]\mbox{\hspace{-#1em}}\right]}

\usepackage{amsmath}

\usepackage{amsfonts}
\usepackage{amssymb}
\usepackage{siunitx} 
\usepackage{cancel}
\usepackage{enumitem}
\usepackage{sectsty}
\usepackage[affil-it,auth-sc]{authblk}
\usepackage{color}
\usepackage[dvipsnames]{xcolor}

\sectionfont{\normalsize}
\subsectionfont{\normalsize}

\usepackage{tikz}
\usetikzlibrary{arrows,decorations,backgrounds,shapes, snakes}
\usetikzlibrary{positioning}
\usetikzlibrary{matrix} 
\usepackage{varwidth}
\usepackage[most]{tcolorbox}
\usetikzlibrary{shapes.geometric,arrows.meta,decorations.markings}
\usepackage{fancyhdr}            
\fancypagestyle{plain}{%
 \fancyhead{}                                     
 \fancyfoot[CE,CO]{\thepage}       
 \fancyhead[CE,CO]{\textcolor[rgb]{0.64,0.15,0.15}{Published in \emph{International Journal of Mechanical Sciences} (2022) \textbf{in press}
 \\
doi: https://doi.org/10.1016/j.ijmecsci.2022.108005}}
\setlength{\headheight}{14.5pt}}

\begin{document}

\newcommand{\singlespace}{\baselineskip=12pt\lineskiplimit=0pt\lineskip=0pt}
\def\ds{\displaystyle}

\tikzstyle{every picture}+=[remember picture]

\newcommand{\beq}{\begin{equation}}
\newcommand{\eeq}{\end{equation}}
\newcommand{\lb}{\label}
\newcommand{\ph}{\phantom}
\newcommand{\beqar}{\begin{eqnarray}}
\newcommand{\eeqar}{\end{eqnarray}}
\newcommand{\barr}{\begin{array}}
\newcommand{\earr}{\end{array}}
\newcommand{\jump}{\parallel}
\newcommand{\Ehat}{\hat{E}}
\newcommand{\That}{\hat{\bf T}}
\newcommand{\Ahat}{\hat{A}}
\newcommand{\chat}{\hat{c}}
\newcommand{\shat}{\hat{s}}
\newcommand{\khat}{\hat{k}}
\newcommand{\muhat}{\hat{\mu}}
\newcommand{\mc}{M^{\scriptscriptstyle C}}
\newcommand{\mei}{M^{\scriptscriptstyle M,EI}}
\newcommand{\mec}{M^{\scriptscriptstyle M,EC}}
\newcommand{\hbeta}{{\hat{\beta}}}
\newcommand{\rec}[2]{\left( #1 #2 \ds{\frac{1}{#1}}\right)}
\newcommand{\rep}[2]{\left( {#1}^2 #2 \ds{\frac{1}{{#1}^2}}\right)}
\newcommand{\derp}[2]{\ds{\frac {\partial #1}{\partial #2}}}
\newcommand{\derpn}[3]{\ds{\frac {\partial^{#3}#1}{\partial #2^{#3}}}}
\newcommand{\dert}[2]{\ds{\frac {d #1}{d #2}}}
\newcommand{\dertn}[3]{\ds{\frac {d^{#3} #1}{d #2^{#3}}}}
\newcommand{\ct}{\captionof{table}}
\newcommand{\cf}{\captionof{figure}}
\newcommand{\sgn}{\text{sgn}}

\def\c{{\circ}}
\def\bob{{\, \underline{\overline{\otimes}} \,}}
\def\ob{{\, \underline{\otimes} \,}}
\def\scalp{\mbox{\boldmath$\, \cdot \, $}}
\def\gdp{\makebox{\raisebox{-.215ex}{$\Box$}\hspace{-.778em}$\times$}}
\def\daa{\makebox{\raisebox{-.050ex}{$-$}\hspace{-.550em}$: ~$}}
\def\mK{\mbox{${\mathcal{K}}$}}
\def\cK{\mbox{${\mathbb {K}}$}}

\def\Xint#1{\mathchoice
   {\XXint\displaystyle\textstyle{#1}}%
   {\XXint\textstyle\scriptstyle{#1}}%
   {\XXint\scriptstyle\scriptscriptstyle{#1}}%
   {\XXint\scriptscriptstyle\scriptscriptstyle{#1}}%
   \!\int}
\def\XXint#1#2#3{{\setbox0=\hbox{$#1{#2#3}{\int}$}
     \vcenter{\hbox{$#2#3$}}\kern-.5\wd0}}
\def\ddashint{\Xint=}
\def\fpint{\Xint=}
\def\dashint{\Xint-}
\def\cpvint{\Xint-}
\def\intl{\int\limits}
\def\cpvintl{\cpvint\limits}
\def\fpintl{\fpint\limits}
\def\ointl{\oint\limits}
\def\bA{{\bf A}}
\def\ba{{\bf a}}
\def\bB{{\bf B}}
\def\bb{{\bf b}}
\def\bc{{\bf c}}
\def\bC{{\bf C}}
\def\bD{{\bf D}}
\def\bE{{\bf E}}
\def\be{{\bf e}}
\def\bbf{{\bf f}}
\def\bF{{\bf F}}
\def\bG{{\bf G}}
\def\bg{{\bf g}}
\def\bi{{\bf i}}
\def\bI{{\bf I}}
\def\bH{{\bf H}}
\def\bK{{\bf K}}
\def\bL{{\bf L}}
\def\bM{{\bf M}}
\def\bN{{\bf N}}
\def\bn{{\bf n}}
\def\bm{{\bf m}}
\def\b0{{\bf 0}}
\def\bo{{\bf o}}
\def\bX{{\bf X}}
\def\bx{{\bf x}}
\def\bP{{\bf P}}
\def\bp{{\bf p}}
\def\bQ{{\bf Q}}
\def\bq{{\bf q}}
\def\bR{{\bf R}}
\def\bS{{\bf S}}
\def\bs{{\bf s}}
\def\bT{{\bf T}}
\def\bt{{\bf t}}
\def\bU{{\bf U}}
\def\bu{{\bf u}}
\def\bv{{\bf v}}
\def\bV{{\bf V}}
\def\bw{{\bf w}}
\def\bW{{\bf W}}
\def\by{{\bf y}}
\def\bz{{\bf z}}
\def\T{{\bf T}}
\def\Te{\textrm{T}}
\def\Id{{\bf I}}
\def\bxi{\mbox{\boldmath${\xi}$}}
\def\balpha{\mbox{\boldmath${\alpha}$}}
\def\bbeta{\mbox{\boldmath${\beta}$}}
\def\bepsilon{\mbox{\boldmath${\epsilon}$}}
\def\bvarepsilon{\mbox{\boldmath${\varepsilon}$}}
\def\bomega{\mbox{\boldmath${\omega}$}}
\def\bphi{\mbox{\boldmath${\varphi}$}}
\def\bsigma{\mbox{\boldmath${\sigma}$}}
\def\bfeta{\mbox{\boldmath${\eta}$}}
\def\bDelta{\mbox{\boldmath${\Delta}$}}
\def\btau{\mbox{\boldmath $\tau$}}
\def\tr{{\rm tr}}
\def\dev{{\rm dev}}
\def\div{{\rm div}}
\def\Div{{\rm Div}}
\def\Grad{{\rm Grad}}
\def\grad{{\rm grad}}
\def\Lin{{\rm Lin}}
\def\Sym{{\rm Sym}}
\def\Skw{{\rm Skew}}
\def\abs{{\rm abs}}
\def\Re{{\rm Re}}
\def\Im{{\rm Im}}
\def\capB{\mbox{\boldmath${\mathsf B}$}}
\def\capC{\mbox{\boldmath${\mathsf C}$}}
\def\capD{\mbox{\boldmath${\mathsf D}$}}
\def\capE{\mbox{\boldmath${\mathsf E}$}}
\def\capG{\mbox{\boldmath${\mathsf G}$}}
\def\tcapG{\tilde{\capG}}
\def\capH{\mbox{\boldmath${\mathsf H}$}}
\def\capK{\mbox{\boldmath${\mathsf K}$}}
\def\capL{\mbox{\boldmath${\mathsf L}$}}
\def\capM{\mbox{\boldmath${\mathsf M}$}}
\def\capR{\mbox{\boldmath${\mathsf R}$}}
\def\capW{\mbox{\boldmath${\mathsf W}$}}

\def\i{\mbox{${\mathrm i}$}}
\def\mC{\mbox{\boldmath${\mathcal C}$}}
\def\mB{\mbox{${\mathcal B}$}}
\def\mE{\mbox{${\mathcal{E}}$}}
\def\mL{\mbox{${\mathcal{L}}$}}
\def\mK{\mbox{${\mathcal{K}}$}}
\def\mV{\mbox{${\mathcal{V}}$}}
\def\C{\mbox{\boldmath${\mathcal C}$}}
\def\E{\mbox{\boldmath${\mathcal E}$}}

\def\AAM{{\it Advances in Applied Mechanics }}
\def\ACME{{\it Arch. Comput. Meth. Engng.}}
\def\ARMA{{\it Arch. Rat. Mech. Analysis}}
\def\AMR{{\it Appl. Mech. Rev.}}
\def\ASCEEM{{\it ASCE J. Eng. Mech.}}
\def\ACTA{{\it Acta Mater.}}
\def\CMAME {{\it Comput. Meth. Appl. Mech. Engrg.}}
\def\CRAS{{\it C. R. Acad. Sci. Paris}}
\def\CRM{{\it Comptes Rendus M\'ecanique}}
\def\EFM{{\it Eng. Fracture Mechanics}}
\def\EJMA{{\it Eur.~J.~Mechanics-A/Solids}}
\def\IJES{{\it Int. J. Eng. Sci.}}
\def\IJF{{\it Int. J. Fracture}}
\def\IJMS{{\it Int. J. Mech. Sci.}}
\def\IJNAMG{{\it Int. J. Numer. Anal. Meth. Geomech.}}
\def\IJP{{\it Int. J. Plasticity}}
\def\IJSS{{\it Int. J. Solids Structures}}
\def\IngA{{\it Ing. Archiv}}
\def\JAM{{\it J. Appl. Mech.}}
\def\JAP{{\it J. Appl. Phys.}}
\def\JAE{{\it J. Aerospace Eng.}}
\def\JE{{\it J. Elasticity}}
\def\JM{{\it J. de M\'ecanique}}
\def\JMPS{{\it J. Mech. Phys. Solids}}
\def\JSV{{\it J. Sound and Vibration}}
\def\MACRO{{\it Macromolecules}}
\def\MMT{{\it Mech. Mach. Th.}}
\def\MOM{{\it Mech. Materials}}
\def\MMS{{\it Math. Mech. Solids}}
\def\MMT{{\it Metall. Mater. Trans. A}}
\def\MPCPS{{\it Math. Proc. Camb. Phil. Soc.}}
\def\MSE{{\it Mater. Sci. Eng.}}
\def\NATURE{{\it Nature}}
\def\NATUREM{{\it Nature Mater.}}
\def\varphiL{{\it Phil. Trans. R. Soc.}}
\def\PMPS{{\it Proc. Math. Phys. Soc.}}
\def\PNAS{{\it Proc. Nat. Acad. Sci.}}
\def\PRE{{\it Phys. Rev. E}}
\def\PRL{{\it Phys. Rev. Letters}}
\def\PRSL{{\it Proc. R. Soc.}}
\def\RIIT{{\it Rozprawy Inzynierskie - Engineering Transactions}}
\def\ROCK{{\it Rock Mech. and Rock Eng.}}
\def\QAM{{\it Quart. Appl. Math.}}
\def\QJMAM{{\it Quart. J. Mech. Appl. Math.}}
\def\SCIENCE{{\it Science}}
\def\SCRMAT{{\it Scripta Mater.}}
\def\SM{{\it Scripta Metall.}}
\def\ZAMM{{\it Z. Angew. Math. Mech.}}
\def\ZAMP{{\it Z. Angew. Math. Phys.}}
\def\ZVDI{{\it Z. Verein. Deut. Ing.}}

\def\salto#1#2{
[\mbox{\hspace{-#1em}}[#2]\mbox{\hspace{-#1em}}]}

\renewcommand\Affilfont{\itshape}
\setlength{\affilsep}{1em}
\renewcommand\Authsep{, }
\renewcommand\Authand{ and }
\renewcommand\Authands{ and }
\setcounter{Maxaffil}{2}

\title{Non-smooth dynamics of buckling based metainterfaces: \\
rocking-like motion and bifurcations
}
\author[1,2]{N. Hima}
\author[3,4]{F. D'Annibale}
\author[1]{F. Dal Corso\footnote{Corresponding author: francesco.dalcorso@unitn.it}}
\affil[1]{DICAM, University of Trento, via~Mesiano~77, I-38123 Trento, Italy}
\affil[2]{FIP MEC srl, via~Scapacchiò~41, 35030 Selvazzano Dentro PD, Italy}
\affil[3]{DICEAA, University of L'Aquila, L'Aquila, Italy}
\affil[4]{M$\&$MOCS, University of L'Aquila, L'Aquila, Italy    }
%

%
%
%
%
%
\maketitle

\begin{abstract}
The non-smooth dynamics is investigated for  an elastic planar metainterface composed by two layers of  buckling elements, each one allowing  motion  on one side only. Through the analogy between buckling and  unilateral contact and by assuming no-bouncing at impact, the motion of the relevant two degrees of freedom system is reduced to that of a single degree governed by a piecewise-smooth differential equation. The  metainterface dynamics has strong similarities with the rocking motion of rigid blocks and displays several types of dynamic bifurcations in the presence of oscillatory forces, including period doubling, branch point cycle, grazing, as well as quasi-periodic and chaotic responses.
Moreover, the multistable response is found to be broaden to conditions representative of  monostable states within a quasi-static setting, disclosing a multistability anticipation by  dynamics. The wide landscape of the dynamic response for the buckling based metainterface provides a novel  theoretical framework to be exploited in  the design of mechanical devices for vibration attenuation and for energy harvesting. 

\end{abstract}



\section{Introduction}
Mechanical instability, a classical scientific field initiated around a quarter of millennium ago by Euler,  is nowadays living a second youth.
Beside the traditional design principles  relied on the rule to avoid  instabilities, a new paradigm emerged  over the last decades:  to exploit  instabilities  for  attaining novel and desired features \cite{kochmann}. This new design perspective is being explored in many technological sectors, encompassing
high-damping and energy-absorbing systems \cite{LEE2012, yap2008, MA2022, MA2022-2}, wave guiding and manipulation \cite{Bertoldi_harnessing_buckl, Guo2018, Xin2016}, deployable structures \cite{ bertoldi_large_def, Melancon, Liu2022} and locomotion \cite{Chen2018, Lee2022-2}.
Towards a further advancement in these sectors,  several intriguing   findings have been recently discovered in the realm of  structural stability as, for example among the many, the buckling under tensile load \cite{Zaccaria2011a, palumbo}, the restabilization of the trivial path for one-dimensional \cite{bosi, BIGONI2014-blade, double_restabilization} and bi-dimensional \cite{FU2019, Healey, bertoldi-spatial} 
structures,  the flutter instability of conservative systems triggered by non-holonomic constraints  \cite{cazzolli, CAZZOLLI2020}, the snapping, coiling, and folding of structures through interaction with fluids \cite{Neukirch2014, Neukirch2018, Neukirch3}, and instability-driven morphologies \cite{XU2020, Xu2022, Siefert2019}.

The modeling and design of meta-structures and meta-interfaces is a challenging task for scholars and researchers in mechanics  towards the  achievement of a designed, tuned,  response from a structure or a material (the reader is refereed to \cite{emilio, Zheludev2012, Zhao2019} 
for mechanical metamaterials, metadevices, and microsystems integration in metamaterials). Interesting phenomena such as complex nonlinear dynamics, bifurcations, chaotic motions may arise and need for an in-depth analytical, numerical and experimental investigation, as the behaviour of many mechanical systems relies on the large amplitude motion of their components. For example, concepts from dynamical system theory are applied to build up a framework for designing fully nonlinear motions of lattice mechanical materials \cite{Jason22}. Complex dynamics can also take place  in nonlinear piezoelectric mechanical systems, where linear and nonlinear damping together with piezoelectric coupling may entail some challenging issues in passive control of flutter instability and post-critical behaviour 
\cite{Danyyy, DAnnibale2016}. 
Moreover, stress and strain distributions  may be controlled 
through active mechanical metamaterials, as for example  by means of piezostack actuators and compliant mechanisms, interconnected to an active metamaterial lattice \cite{saravana22}. 
Advanced dynamic performances can be also attained through specific  periodic microstructures realized with, among different topologies,   one-dimensional cellular lattices build up by a pantograph mechanism in the tetra-atomic cell and behaving as an inertially amplified metamaterial \cite{SETTIMI2021}. 

\begin{figure}[!ht]
	\centering
	\includegraphics[width=\textwidth]{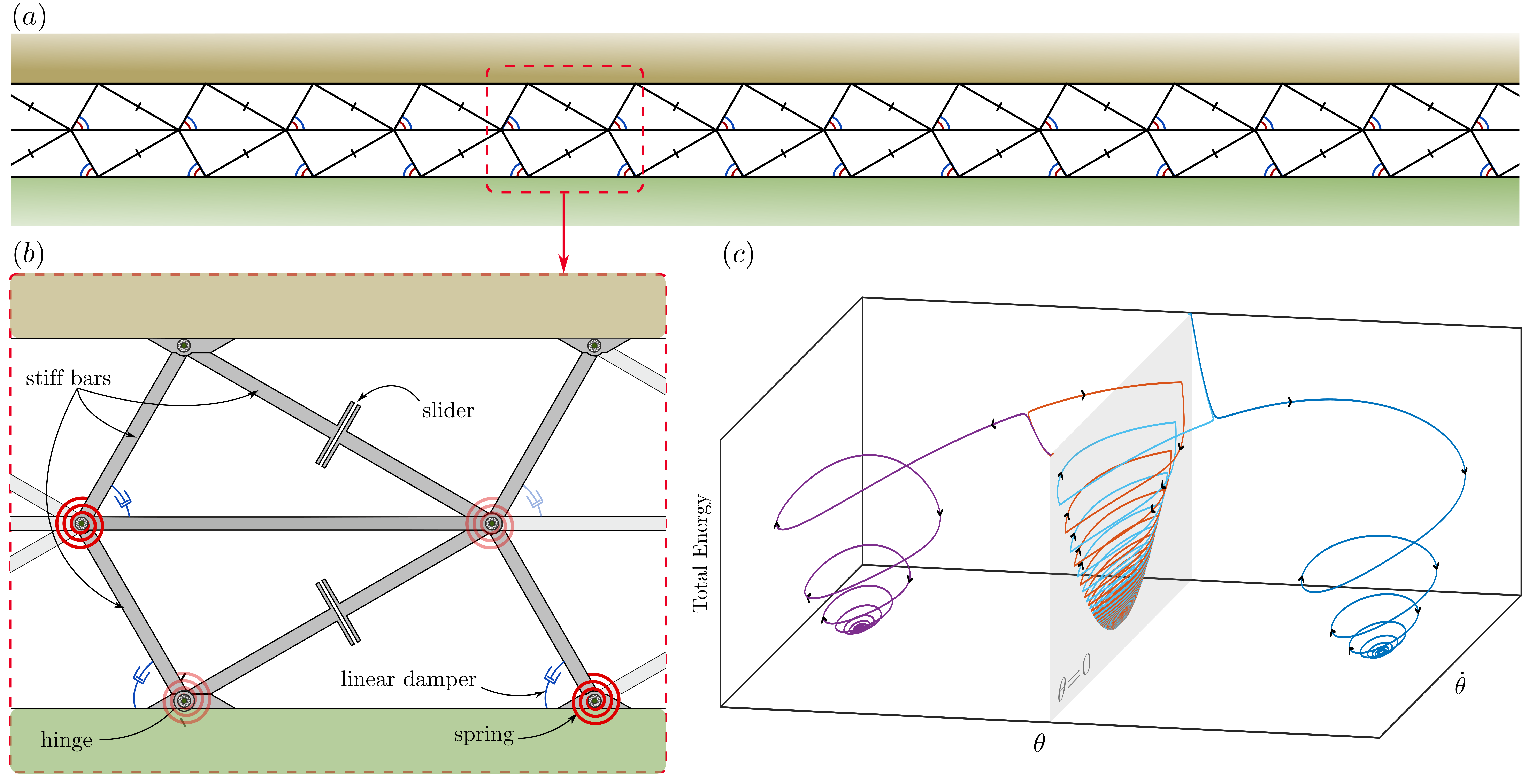}
	\caption{$(a)$ The metainterface based on two layers of tensile-buckling elements and displaying  a piecewise-smooth dynamics. $(b)$ The unit structural cell composed by stiff bars, rotational springs, dampers and sliders. $(c)$  Vibrations trajectories  for  a nonconservative tristable metainterface  under  constant symmetric loading conditions  (dataset of Fig. \ref{phase-diagram-sym}-$c$) within the rotation-angular velocity-total energy space, together with the system discontinuity plane (grey) corresponding to the undeformed configuration.}
	\label{intro_pic}
\end{figure}

 Within this context, the mechanical response of an elastic metainterface (Fig. \ref{intro_pic}-$a$) based on two layers of buckling elements is investigated. The unit structural cell of this interface (Fig. \ref{intro_pic}-$b$) is composed by stiff bars joined together through hinges and it can be deformed by storing  the  elastic energy in two rotational springs and by means of the slider presence along two diagonal bars. In particular, the slider enables the bifurcation of the relevant bar at null tensile axial force, which in turn may be exploited to realize a desired mechanical behaviour.
 The quasi-static behaviour  of this unit element has been analyzed in \cite{hima22} by showing (i.) the equivalence of the tensile (or compressive) buckling element with a unilateral constraint and (ii.) the possibility to realize a  multistable device with tunable mechanical response, despite the corresponding small number of degrees of freedom.

The objective of the present research is to extend  the mechanical behaviour analysis of the unit cell to dynamic conditions. The kinematics of this system is dependent on  two degree of freedom system (Sect. \ref{kinsec}) and the motion is governed by two coupled nonlinear ordinary differential equations subject to unilateral constraints (Sect. \ref{lagsec}). Such constraints limit each layer of the structure to a unilateral motion, which in turn
leads to various impact scenarios, differing in the bouncing, no-bouncing, or partial-bouncing of the impacting layer (Sect. \ref{impsec}).
Under the assumption of no bouncing occurrence, the motion analysis reduces to that of a single degree of freedom described by a  piecewise-smooth differential equation \cite{simson-book, ZHANG20221, CHARROYER201890}, with the switching plane corresponding to the undeformed equilibrium configuration (Fig. \ref{intro_pic}-c). The non-smoothness in the response is always realized through an acceleration jump and, depending on impact dissipation phenomena, also through a possible velocity jump.
 
Vibration analysis under constant symmetric loading reveals strong similarities with the classical rocking motion of rectangular rigid bodies in rigid surfaces (Sect. \ref{const_load_vibr}), mainly differing for the  absence of the first order term of the rotations due to rotational spring in the present system. Small amplitude vibration analysis sheds light over the finite time decay of the free rocking-like motion (Sect. \ref{small_sec}). 
Phase portraits at large amplitude vibrations disclose the influence of  different damping mechanisms and system non-symmetry on the motion, showing the possible existence of three stable attractors within a feasible range of deformations (Sect. \ref{phase_sec}). Moreover, systems with extreme stiffness ratios display  an intriguing bouncing-like behaviour even under no bouncing assumption (Sect. \ref{bounc_sec}).
 
Response spectra and bifurcation diagrams reveal a rich mechanical behaviour of the system under oscillatory loading conditions, including primary and super-harmonic resonances, period-doubling cascades,  quasi-periodic and chaotic behaviour, grazing, and coexistence of  stable equilibrium paths within a chaotic region (Sect \ref{forcedvib}). 
Finally, depending on the force velocity, the interplay between external and parametric excitation  may provide a bistable response at load amplitudes corresponding to a monostable response under the quasi-static assumption. As a consequence, dynamic effects realize a multistability anticipation for the system.
 
The obtained results show the possibility to harness bifurcation phenomena of the present structural system  to display highly desirable mechanical features, without compromising the global stability of the system. Moreover, the multistable and multisource dissipation mechanisms embedded in the considered metainterface concept  open new possibilities in the design of technological devices for vibration attenuation and  energy harvesting.
 
\section{Mechanical model for the structural unit cell}\label{kinsec}
\begin{figure}[!ht]
	\centering
	\includegraphics[width=\textwidth]{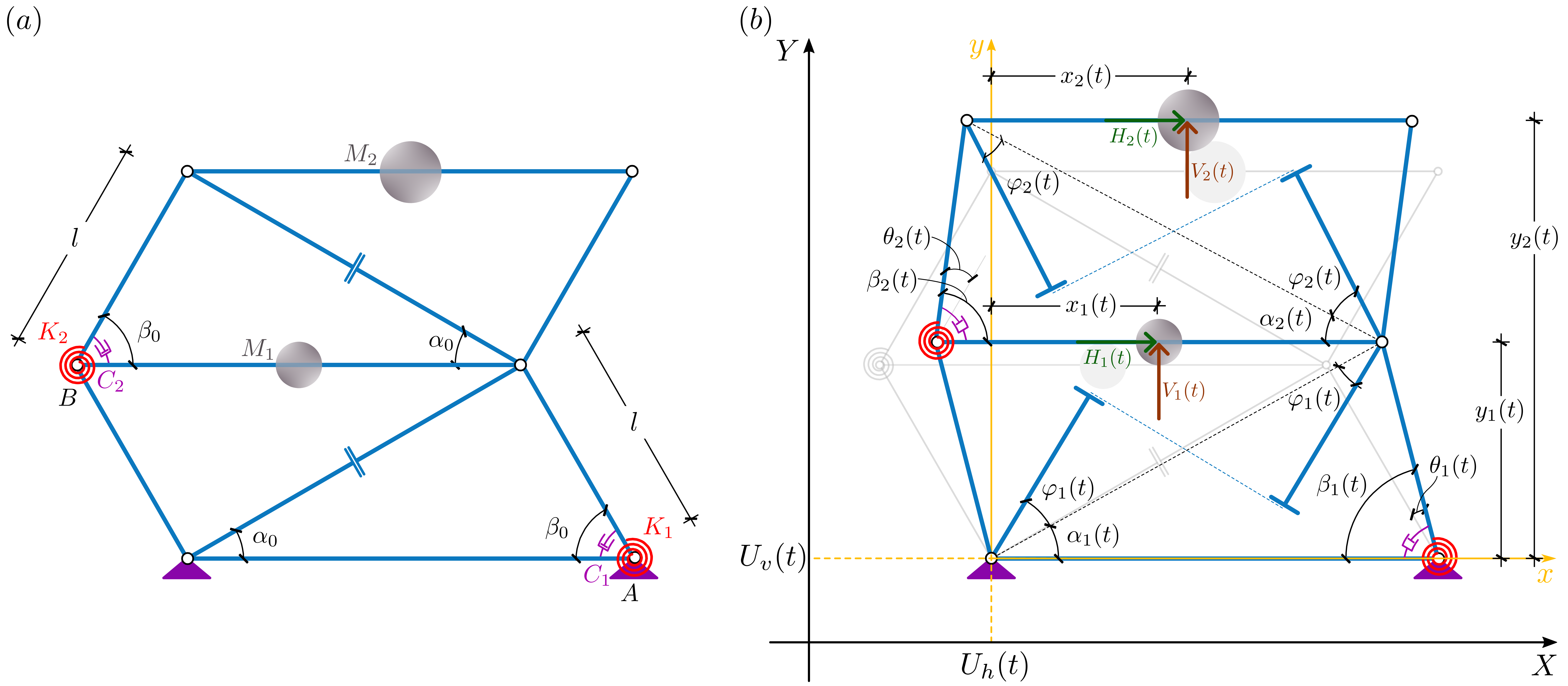}
	\caption{($a$) Undeformed  and ($b$) deformed configuration for the unit structural cell composing the metainterface in Fig. \ref{intro_pic}. The unit cell is made up of two  articulated quadrilateral structures, each one composed of rigid bars and containing a visco-elastic hinge (of stiffness $K_j$ and viscous coefficient $C_j$, $j=1,2$) and a slider at the midpoint of the diagonal bar within the parallelogram. Loading is applied by means of horizontal $H_j(t)$ and vertical $V_j(t)$ forces at the $j$-th layer and of ground  horizontal $U_h(t)$ and vertical $U_v(t)$ displacements. The deformed configuration can be described through the misalignment angles  $\varphi_j$ or through the  difference angles $\theta_j$, where the latter measure is the most  appropriate to investigate the dynamic response.}
	\label{intro}
\end{figure}

The mechanical response is addressed for a planar structural system made up of two layers of an articulated quadrilateral structure with parallelogram shape, realized by rigid bars connected to one another through hinges  and with dimension scaled through the length $l$, corresponding to that of the inclined bars along the parallelogram perimeter (Fig. \ref{intro}). 
Two visco-elastic hinges  are present by adding  linear rotational springs of stiffness $K_1$ and $K_2$, and  dashpots of viscous coefficients $C_1$ and $C_2$ respectively to the hinges located at the points  \lq A' and \lq B', while two sliders are located at the mid-span of a diagonal of each parallelogram.
Since each slider constraints the continuity in axial and rotational displacements but allows for a jump in the transverse displacement, the $j$-th layer may deform into another parallelogram described by the configuration angles $\alpha_j$ and $\beta_j$, which are functions of the misalignment angle $\varphi_j$ displayed at the relevant slider as
\begin{equation}\label{gen_conf}
\alpha_j(\varphi_j) = \arccos\left(\dfrac{ \psi}{2 \lambda \cos \varphi_j} + \chi \cos \varphi_j\right),\qquad\beta_j(\varphi_j) = \arccos\left(\cos \beta _0 - \psi \tan ^2\varphi_j \right)\geq \beta_0,
\end{equation}
where $\psi$, $\lambda$ and $\chi$ are constants defined as 
\begin{equation}\label{psi_lambda_chi}
    \psi = \frac{\sin \beta_0^2}{2 \sin \alpha_0  \sin (\alpha _0+\beta _0)}, \qquad \lambda = \dfrac{\sin \beta_0}{2 \sin \alpha_0}>0, \qquad 
    \chi = \dfrac{ \sin (2\alpha_0 + \beta_0)}{\sin (\alpha_0 + \beta_0)},
\end{equation}
being  $\alpha_0$ and $\beta_0$  the angles describing the undeformed configuration, $\alpha_j(\phi_j=0)=\alpha_0$
and $\beta_j(\phi_j=0)=\beta_0$, and subject to the following constraints
\begin{equation}\label{restangle}
	\alpha_0>0,\qquad
	0<\beta_0<\pi,\qquad
	\alpha_0+\beta_0<\pi.
\end{equation}

The inertia of the system is modelled through the mass $M_j$ attached to the central point along the top edge of the $j$-th layer. Because the motion of the top edge of each layer is described by a rigid translation without any rotation, the possible presence of rotational inertia has no effect in the present analysis. The position of the mass relevant to the $j$-th layer is described with varying of the physical time $t$ through its absolute coordinates $X_j(t)$ and $Y_j(t)$, evaluated as
\begin{equation}\label{eq:1}
	\begin{cases}
	X_j(t)=U_h(t)+x_j(t),\\
	Y_j(t)=U_v(t)+y_j(t),
	\end{cases}
\end{equation}
where $x_j(t)$ and $y_j(t)$ are the relative coordinates of the same point measured in the  $x-y$ non-inertial frame, attached to the lattice base and with origin located at the lower left hinge, while $U_{h}(t)$ and $U_{v}(t)$ measure the ground motion of the $x-y$ non-inertial frame with respect to the   $X-Y$ inertial frame.

In order to proceed with a non-dimensional analysis, the dimensionless time $\tau$  is introduced as
\beq
\tau=\frac{t}{T},
\eeq
where the characteristic time $T$ is considered as
\begin{equation}\label{timeT}
	T=\sqrt{\frac{M_2l^2}{K_2}}.
\end{equation}
Moreover, dimensionless ground motion components  are introduced as
\begin{equation}\label{displ_ratio1}
	\zeta(\tau)=\frac{U_h(\tau)}{l},\qquad
	\nu(\tau)=\frac{U_v(\tau)}{l},
\end{equation}
as well as  the following relative dimensionless horizontal and vertical coordinates 
\begin{equation}\label{displ_ratio2}
	\begin{array}{cc}	
		\xi_1(\tau)=\dfrac{x_1(\tau)}{ l}+\dfrac{\sin \left(\alpha _0-\beta _0\right)}{2 \sin \alpha_0},\qquad
		\xi_2(\tau)=\dfrac{x_2(\tau)-x_1(\tau)}{ l} - \cos\beta_0,\\[6mm]
		\eta_1(\tau)=\dfrac{y_1(\tau)}{ l},\qquad
		\eta_2(\tau)=\dfrac{y_2(\tau)-y_1(\tau)}{ l}.
	\end{array}
\end{equation}

\begin{figure}[!ht]
	\centering
	\includegraphics[width=\textwidth]{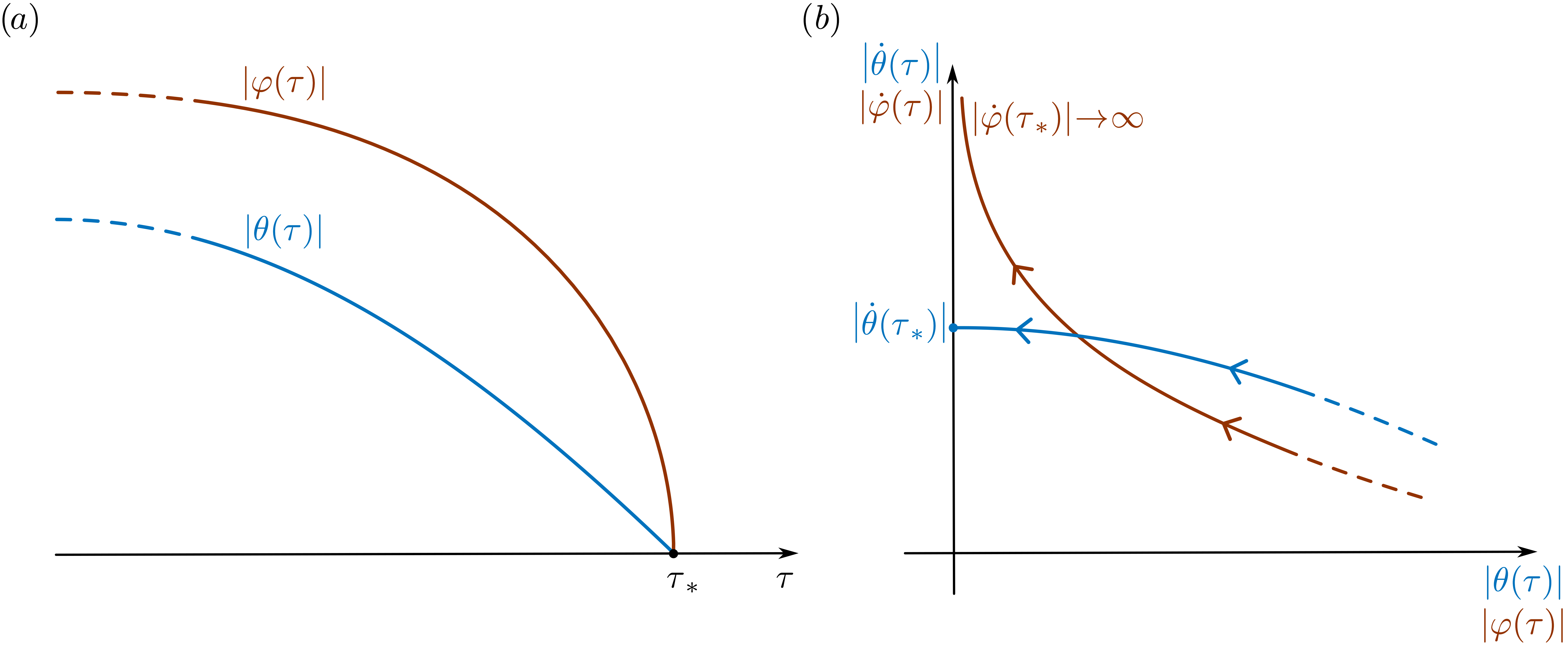}
	\caption{Schematic representation of $(a)$ the evolution in the dimensionless time $\tau$ and  of $(b)$ the phase portraits for the modulus of the misalignment angle  $\left|\varphi\right|$ and of the difference angle $\left|\theta\right|$ just before the impact  time $\tau_*$.  A singular value for the velocity of the misalignment angle is displayed at the impact, $|\dot\varphi(\tau_*)| \rightarrow \infty$,  therefore the dynamic analysis of the structural system has to be performed with reference to the difference angle $\theta(\tau)$ in order to overcome numerical integration issues. }
	\label{singularity}
\end{figure}

Interestingly, the deformed configuration of the $j$-th layer can also be described through the difference angle $\theta_j$ at the respective visco-elastic hinge, defined as a function of the misalignment angle $\varphi_j$ through 
\beq
\theta_j(\tau)=(-1)^j\left[\beta_j(\varphi_j(\tau))-\beta_0\right],
\eeq
implying  the following inequalities
\begin{equation}\label{theta_conditions}
	\theta_1(\tau)\geq 0,\qquad
	\theta_2(\tau)\leq 0\qquad\Rightarrow\qquad
	\theta_1(\tau)\theta_2(\tau)\leq 0.
\end{equation}
In the quasi-static analysis of the present system \cite{hima22}, it has been shown that the deformed configuration of the layers is realized through a bifurcation of system referred to the misalignment angles $\varphi_j$  or in the presence of unilateral constraints  when the difference angles $\theta_j$ are considered, because restricted to satisfy eqn. (\ref{theta_conditions}).
Although in practical terms the two measures of angles are  equivalent  within a quasi-static setting, the difference angles $\theta_j(\tau)$ represent the best measure to refer when performing a  dynamic  analysis. Indeed,  the misalignment angle $\varphi_j$ assumes infinite  velocity at the impact of the $j$-th layer, namely when it  approaches zero, providing difficulties in the convergence of numerical integration of the equation of motions at each impact condition. This is made evident by recalling that  the misalignment and the difference angles are related to each other under small rotations through
\begin{equation}\label{fundamentalproperty}
\left|\theta_j(\tau)\right|\approx \frac{\psi}{\sin \beta_0} \, \varphi_j^2 (\tau), \qquad
\mbox{for} \qquad \left|\varphi_j\right|\rightarrow 0,
\end{equation}
and evaluating its derivative,
\begin{equation}\label{fundamentalvelocity}
\left|\dot\theta_j(\tau)\right|\approx \frac{2\psi}{\sin \beta_0} \, \left|\varphi_j(\tau)\right| \left|\dot \varphi_j(\tau)\right|,
\end{equation}
the singular behaviour of the misalignment angle velocity $ \dot \varphi_j(\tau^*)$ at the impact time $\tau^*$ (defining  $\varphi_j(\tau^*)=0$) is implied by assuming  a corresponding finite magnitude of the  difference angle velocity $\left|\dot\theta(\tau^*)\right|$. This behaviour is  graphically represented in Fig. \ref{singularity} through $a$)  the generic evolution in time of the angle measures and $b$) their  phase portraits  are reported for time $\tau$ close to that of impact $\tau^*$.

Due to the above-mentioned motivation, it is instrumental to express the kinematic parameters $\xi_j$, and $\eta_j$  to the respective difference angle $\theta_j$ through the following expressions \footnote{For completeness, the  configuration angle  $\alpha_j$ as a function of the difference angle $\theta_j$ is given by
\begin{equation}\label{configalpha}
	\alpha_j(\tau)=\ds\text{arcsec}\left[\sqrt{\frac{\cos \beta_0 - \cos \left(\beta _0+(-1)^{j} \theta _j(\tau )\right)+\psi }{\psi }}\frac{2 \lambda }{\cos \beta_0 - \cos \left(\beta _0+(-1)^{j} \theta _j(\tau )\right)+2 \lambda  \chi +\psi }\right].
\end{equation}}

\begin{equation}\label{configdbeta}
	\xi_j(\tau)=-(-1)^j \left(\cos \beta _0-\cos \left(\beta _0-(-1)^j \theta _j(\tau )\right)\right), \qquad
	\eta_j(\tau)=\sin\left(\beta_0-(-1)^j \theta_j(\tau)\right).
\end{equation}

Finally, it is noted that  the $j$-th parallelogram reduces into a line segment when the difference angle takes the following special values
\begin{equation}\label{line_cond}
\left|\theta_j\right|=\overline{\theta}^{[n]},\qquad
\mbox{where}\qquad
\overline{\theta}^{[n]}=n\pi-\beta_0,\qquad n\in\mathbb{N}.
\end{equation}
The above condition evaluated at $n=1$ may represent a  
limitation to the  motion of the system if impact between the different rigid bars occurs. However, if the related dimension of connected object allows, this situation can be circumvented by realizing the structural system  exploiting the out-of-plane direction. Indeed, when the  rigid bars are located on different planes, their rotations can monotonically increase in magnitude without any  limitation except for the energy that can be possibly stored within the respective rotational spring.
In any case, independently of this aspect, eqn. \eqref{line_cond} defines an upper bound to the magnitude of the relative horizontal displacement $\xi_j$ that can be displayed by the mass $M_j$
\begin{equation}\label{max_disp_condition}
	|\xi_j|<\overline{\xi}, \qquad \mbox{where}\qquad
	\overline{\xi}=1+\cos\beta_0.
\end{equation}

\section{The Lagrangian and the  equations of motion}\label{lagsec}

The kinetic energy $\mathcal{T}$ of the system under consideration (Fig. \ref{intro}) can be expressed as a function of the difference angles $\theta_j$ and their velocities $\dot\theta_j$ ($j=1,2$) as
\begin{equation}\label{kin_eng}
	\begin{array}{lll}
		\ds \mathcal{T}  = \ds\frac{K_2}{2}\sum_{j=1}^{2}& [(2-j)m+j-1] \left\{ \left[\dot{\zeta}(\tau) + (j-1) \sin (\beta_0 + \theta_1(\tau) ) \dot{\theta}_{1}(\tau)  + \sin \left(\beta_0 - (-1)^j \theta_j(\tau)\right) \dot{\theta}_{j}(\tau)\right]^2\right.\\[4mm]
		&\left. + \left[\dot{\nu}(\tau) + (j-1) \cos (\beta_0 + \theta_1(\tau) ) \dot{\theta}_{1}(\tau) - (-1)^j \cos (\beta_0 - (-1)^j \theta_j (\tau) ) \dot{\theta}_{j}(\tau) \right]^2 \right\},
	\end{array}
\end{equation}
while its potential energy $\Pi$ is given as the sum of the elastic energy stored in the two rotational springs as follows

\begin{equation}\label{pot_eng_beta}
	\begin{array}{lll}
		\Pi = \ds\frac{K_2 }{2}\sum_{j=1}^{2}\ds[(2-j) k+j-1]\theta_j(\tau)^2,
	\end{array}
\end{equation}
with the mass ratio $m$ and the stiffness ratio $k$ introduced as  
\begin{equation}\label{m_k_ratio}
	m=\dfrac{M_1}{M_2}, \qquad k=\dfrac{K_1}{K_2}.
\end{equation}

The presence of the viscous dampers (with damping coefficient $C_j$) at the hinges located at points \lq A' and \lq B' (Fig. \ref{intro}), to account for the possible air resistance and other sources of dissipation during the motion, leads
to the following Rayleigh dissipation function $\mathcal{R}$ 
\begin{equation}\label{Rayleigh}
	\begin{array}{lll}
		\mathcal{R} = \ds K_2 \frac{\mathcal{C} }{2}\sum_{j=1}^{2}\ds[(2-j) c+j-1]\dot{\theta}_j(\tau)^2,
	\end{array}
\end{equation}
where the dissipation ratio $c$ and the damping ratio $\mathcal{C}$ are introduced as
\begin{equation}\label{disip_ratio}
	c=\dfrac{C_1}{C_2},\qquad
	\mathcal{C}=\dfrac{C_2}{M_2 l^2}.
\end{equation}

By considering the Lagrangian $\mathcal{L}=\mathcal{T}-\Pi$ and the Rayleigh dissipation function $\mathcal{R}$, the equations of motion can be obtained   through the Euler-Lagrange equation,

\begin{equation}\label{dyn_eqn_beta}
	\dfrac{\mbox{d}}{\mbox{d} \tau}\left(\dfrac{\partial \mathcal{L}\left(\dot{\theta_1}, \dot{\theta_2},\theta_1, \theta_2\right)}{\partial \dot{\theta_j}}\right)-\dfrac{\partial\mathcal{L}\left(\dot{\theta_1},\dot{\theta_2},\theta_1, \theta_2\right)}{\partial  \theta_j} + \dfrac{\partial\mathcal{R}\left(\dot{\theta_1},\dot{\theta_2}\right)}{\partial  \dot{\theta}_j}=\mathcal{Q}_{\theta_j}(\tau),\qquad
	{j=1,2},
\end{equation}

where $\mathcal{Q}_{\theta_j}(\tau)$ is the generalized  external forces evaluated through  the principle of the virtual work for a virtual rotation $\delta\theta_j$. By assuming  the set  of horizontal $H_1(t)$, $H_2(t)$ and vertical $V_1(t)$, $V_2(t)$  forces acting on the masses $M_1$ and $M_2$ (as sketched in Fig. \ref{intro}), the generalized forces $\mathcal{Q}_{\theta_j}(\tau)$ are evaluated as 
\begin{equation}\label{ext_forces_beta}
	\begin{array}{ll}
	\mathcal{Q}_{\theta_j}(\tau) = & K_2 \left[  \left(1+h(\tau)\right)^{2-j}\mathcal{H}(\tau)\sin(\beta_0-(-1)^j \theta_j(\tau))\right.\\[3mm]
		 & \left.  - (-1)^j\left(1+v(\tau)\right)^{2-j} \mathcal{V}(\tau) \cos (\beta_0 - (-1)^j \theta_j(\tau) )\right],
	\end{array}\qquad\qquad
	{j = 1, 2},
\end{equation}

where the dimensionless external horizontal $\mathcal{H}(\tau)$ and vertical $\mathcal{V}(\tau)$ forces  and their respective ratios, $h(\tau)$ and $v(\tau)$, have been introduced as
\begin{equation}\label{dim_less}
	\begin{array}{cc}
		 \mathcal{H}(\tau)=\dfrac{H_2(\tau)l}{K_2}, \qquad \mathcal{V}(\tau)=\dfrac{V(\tau)l}{K_2}, \qquad
		h(\tau) = \dfrac{H_1(\tau)}{H_2(\tau)}, \qquad v(\tau) = \dfrac{V_1(\tau)}{V_2(\tau)}.	
	\end{array}
\end{equation}

The  equations of motion are then given by the following system of two coupled nonlinear ordinary differential equations for  the difference angles $\theta_j(\tau)$, restricted by the inequalities (\ref{theta_conditions}),
 \begin{equation}\label{eomnl}
 \begin{array}{rl}
     \mathbf{M}\left(\theta _1(\tau ),\theta _2(\tau )\right)
     \begin{Bmatrix}
		    \ddot{\theta}_1(\tau) \\[2mm]
			\ddot{\theta}_2(\tau)
		\end{Bmatrix}
		+
		\mathbf{C} \begin{Bmatrix}
			\dot{\theta}_1(\tau) \\[2mm]
		    \dot{\theta}_2(\tau)
		\end{Bmatrix}
		&	+
			\mathbf{D}\left(\theta _1(\tau ),\theta _2(\tau )\right) \begin{Bmatrix}
			\dot{\theta}_1^2(\tau) \\[2mm]
		    \dot{\theta}_2^2(\tau)
		\end{Bmatrix}
		\\[6mm]&+
			\mathbf{K}\left(\theta _1(\tau ),\theta _2(\tau )\right) \begin{Bmatrix}
			\theta_1(\tau) \\[2mm]
			\theta_2(\tau)
		\end{Bmatrix}
		=\mathbf{f}\left(\theta _1(\tau ),\theta _2(\tau )\right),
		\end{array}
 \end{equation}
 where $ \mathbf{M}$ is the (symmetric) mass operator, $\mathbf{C}$ is the diagonal viscous damping matrix, $\mathbf{D}$ is the (skew-symmetric) operator multiplying the squared angular velocities (springing from kinetic nonlinearities),  $\mathbf{K}$ is the diagonal stiffness  operator, and
 $\mathbf{f}$ is the equivalent force vector, defined as
 \begin{equation}
	\begin{array}{ll}
		\mathbf{M}=\begin{bmatrix}
			m+1 & -\cos \left(2 \beta _0 + \theta _1(\tau )-\theta _2(\tau )\right) \\[2mm]
			-\cos \left(2 \beta _0+\theta _1(\tau )-\theta _2(\tau )\right) & 1
		\end{bmatrix},\qquad
		 \mathbf{C}=
	\mathcal{C}	\begin{bmatrix}
			c\;  & 0 \\[2mm]
			0 & 1
		\end{bmatrix},\\[6mm]
\mathbf{D}=			\begin{bmatrix}
			0 & -\sin \left(2\beta_0+\theta _1(\tau )-\theta _2(\tau )\right) \\[2mm]
			\sin \left(2 \beta_0+\theta_1(\tau) - \theta_2(\tau)\right) & 0
		\end{bmatrix},\qquad
\mathbf{K}=	
		\begin{bmatrix}
			k\;  & 0 \\[2mm]
			0 & 1
		\end{bmatrix},\\[6mm]
		\mathbf{f}=	\begin{Bmatrix}
			\left[\left(1+h(\tau)\right)\mathcal{H}(\tau) - (m+1) \ddot{\zeta}(\tau ) \right] \sin (\beta_0+\theta_1(\tau)) + \left[\left(1+v(\tau)\right)\mathcal{V}(\tau) - (m+1) \ddot{\nu}(\tau) \right] \cos \left(\beta_0+\theta_1(\tau)\right) \\[2mm]
			\left[\mathcal{H}(\tau) - \ddot{\zeta}(\tau ) \right] \sin (\beta_0-\theta_2(\tau)) - \left[\mathcal{V}(\tau) - \ddot{\nu}(\tau) \right] \cos (\beta_0 - \theta_2(\tau))
		\end{Bmatrix}.
	\end{array}
\end{equation}

Under the assumption of oscillations with small amplitudes  $\theta_j(\tau)$, the equations of motion \eqref{eomnl} can be rewritten  as the first order truncation of  Taylor series  as
\begin{equation}
     \mathbf{M}_0
     \begin{Bmatrix}
		    \ddot{\theta}_1(\tau) \\[2mm]
			\ddot{\theta}_2(\tau)
		\end{Bmatrix}
		+
		\mathbf{C} \begin{Bmatrix}
			\dot{\theta}_1(\tau) \\[2mm]
		    \dot{\theta}_2(\tau)
		\end{Bmatrix}
				+
			\mathbf{K}_0(\tau ) \begin{Bmatrix}
			\theta_1(\tau) \\[2mm]
			\theta_2(\tau)
		\end{Bmatrix}
		=\mathbf{f}_0(\tau),
 \end{equation}
 where
 \begin{equation}
	\begin{array}{ll}
		\mathbf{M}_0=\begin{bmatrix}
			m+1 & -\cos 2\beta_0 \\[2mm]
			-\cos 2\beta_0 & 1
		\end{bmatrix},\qquad
		\mathbf{K}_0=	
		\begin{bmatrix}
			\kappa_1(\tau) & 0 \\[2mm]
			0 & \kappa_2(\tau )
		\end{bmatrix},\\[6mm]
		\mathbf{f}_0=	\begin{Bmatrix}
			\sin \beta_0 \left[\left(h(\tau)+1\right) \mathcal{H}(\tau)-(m+1) \ddot{\zeta}(\tau) \right]+\cos \beta_0 \left[\left(v(\tau)+1\right) \mathcal{V}(\tau)-(m+1) \ddot{\nu}(\tau)\right] \\[2mm]
			\sin \beta_0 \left[\mathcal{H}(\tau )-\ddot{\zeta}(\tau)\right]-\cos \beta _0 \left[\mathcal{V}(\tau )-\ddot{\nu}(\tau)\right]
		\end{Bmatrix},
	\end{array}
\end{equation}
with 
\begin{equation}
	\begin{array}{ll}
		\kappa_j(\tau ) =& k^{2-j} + (-1)^j \cos \beta_0 \left[\left(h(\tau)+1\right)^{2-j} \mathcal{H}(\tau )-(m+1)^{2-j} \ddot{\zeta}(\tau )\right]\\[2mm]
		&+\sin \beta_0 \left[\left(v(\tau) + 1\right)^{2-j} \mathcal{V}(\tau)-(m+1)^{2-j}\ddot{\nu}(\tau )\right].
	\end{array}
\end{equation}
It is noted that both the equivalent force vector $\mathbf{f}_0$ and  the stiffness operator\footnote{It is noted that, for simplicity in nomenclature, $\mathbf{K}_0$ is defined as the difference of the expansion of $\mathbf{K}$ and the coefficients of  the linear terms in $\theta_j$ of the expansion of $\mathbf{f}$.} $\mathbf{K}_0$ have non-constant coefficients whenever the external forces or the ground accelerations vary in time, providing a simultaneous external and parametric excitation to the system.

The obtained equations of motion are complemented in the next Section by  the analysis of the possible  impact scenarios introduced by the unilateral  constraints  (\ref{theta_conditions}) on the difference angles $\theta_j(\tau)$.

\section{Impact scenarios, no bouncing assumption, and reduction to a single degree of freedom}\label{impsec}
During the motion, the deformed $s$-th layer of the planar structure may return to its undeformed configuration at the time $\tau_*$, $\theta_s(\tau_*)=0$. Because the $s$-th layer can not smoothly continue its motion at $\tau_*$ due to the unilateral constraint on the difference angle $\theta_s$, eqn. (\ref{theta_conditions}), an impact occurs realizing an impulse to the structure. The effect of such an impulse is modelled as a non-smooth (instantaneous) transition expressed by a possible jump\footnote{In the following, the brackets $\salto{0.38}{\, \cdot \, }$ indicate the jump in the corresponding quantity (velocity or acceleration) at the impact moment $\tau_*$ and is defined as the difference of that quantity after and before the impact
\beq\label{jump_function}
	\salto{0.38}{g(\tau_*) } = g(\tau_*^+)-g(\tau_*^-).
\eeq} in the layers velocities  at the impact time $\tau_*$, while the rotations are continuous
\beq\label{controtrot}
	\theta_n(\tau_*^-)=\theta_n(\tau_*^+),\qquad
	n=1,2.
\eeq
Considering the $s$-th layer as the impacting one while the $j$-th layer as the other one ($j\neq s$), the possible impact scenarios are listed in Table \ref{tab:impact_scenarios} in terms of the difference angles of the layers just before ($\tau_*^-$) and just after ($\tau_*^+$) the impact. While two cases are possible before the impact (depending on the number of deformed layers), after the impact three conditions can be realized, namely
\begin{itemize}
    \item  \emph{no bouncing}, for which the impacting layer stops and the other one has a velocity jump
\begin{equation}\label{impact_condition1}
		\dot\theta_s(\tau_*^-)\dot\theta_s(\tau_*^+)= 0,
		\qquad
		\dot\theta_j(\tau_*^-)\neq\dot\theta_j(\tau_*^+),
\end{equation}
\item \emph{bouncing}, for which the impacting layer reverses its motion and the other one has no velocity jump
\begin{equation}\label{impact_condition2}
		\dot\theta_s(\tau_*^-)\dot\theta_s(\tau_*^+)< 0,\qquad
		\dot\theta_j(\tau_*^-)=\dot\theta_j(\tau_*^+),
\end{equation}
\item  \emph{partial bouncing}, for which the impacting layer reverses its motion and the other one has a velocity jump
\begin{equation}\label{impact_condition3}
		\dot\theta_s(\tau_*^-)\dot\theta_s(\tau_*^+)< 0,\qquad
		\dot\theta_j(\tau_*^-)\neq\dot\theta_j(\tau_*^+).
\end{equation}

\end{itemize}

\begin{table}[!ht]
\caption{Impact scenarios of the planar lattice when the $s$-th layer is impacting at the time $\tau_*$ ($s,j=1,2$ with $s\neq j$)}
\label{tab:impact_scenarios}
\begin{center}
\begin{tabular}{|cl|cll|}
\hline
\multicolumn{2}{|c|}{Just before impact ($\tau=\tau_*^{-}$)}     & \multicolumn{3}{c|}{Just after impact  ($\tau=\tau_*^{+}$)}         \\[3pt] \hline
\multicolumn{1}{|c|}{    
    $\begin{array}{cc}
    \# \,\mbox{Deformed}\\
    \mbox{layers}
    \end{array}$ }                  &        Condition      & \multicolumn{1}{c|}{
    $\begin{array}{cc}
    \# \,\mbox{Deformed}\\
    \mbox{layers}
    \end{array}$}             & \multicolumn{2}{c|}{Condition} \\[3pt] \hline
\multicolumn{1}{|c|}{\multirow{3}{*}{
$\begin{array}{c}
    \\
    \\
    1
    \end{array}$
}} & \multirow{3}{*}{
$\begin{array}{ll}
    \\[4pt]
    \theta_s(\tau_*^-)\rightarrow 0,\\[2mm]
    \theta_j(\tau_*^-)=0
    \end{array}$
    } & \multicolumn{1}{c|}{\multirow{2}{*}{
    $\begin{array}{c}
    \\
    1
    \end{array}$
    }} &         
    $\begin{array}{ll}
    \theta_s(\tau_*^+)=0,\\
    \theta_j(\tau_*^+)\neq 0
    \end{array}$
    &   \emph{No Bouncing}        \\[3pt] \cline{4-5} 
\multicolumn{1}{|c|}{}                  &                   & \multicolumn{1}{c|}{}  & 
    $\begin{array}{ll}
     \theta_s(\tau_*^+)\neq0,\\
     \theta_j(\tau_*^+)=0
     \end{array} $
& \emph{Bouncing}\\ \cline{3-5} 
\multicolumn{1}{|c|}{}                  &                   & \multicolumn{1}{c|}{2} &  
    $\begin{array}{ll}
     \theta_s(\tau_*^+)\neq 0,\\
     \theta_j(\tau_*^+)\neq 0
     \end{array} $
& \emph{Partial Bouncing} \\[3pt] \hline
\multicolumn{1}{|c|}{\multirow{2}{*}{2}}& \multirow{2}{*}{
    $\begin{array}{ll}
     \theta_s(\tau_*^-)\rightarrow 0,\\[2mm]
     \left|\theta_j(\tau_*^-)\right|=\theta_*>\left|\theta_s(\tau_*^-)\right|
     \end{array} $}
     & \multicolumn{1}{c|}{1}                  &  
    $\begin{array}{ll}
    \theta_s(\tau_*^+)=0,\\
    \theta_j(\tau_*^+)\neq \theta_*
    \end{array}$ 
     &  \emph{No Bouncing} \\[3pt] \cline{3-5} 
\multicolumn{1}{|c|}{}                  &                   & \multicolumn{1}{c|}{2} &    
    $\begin{array}{ll}
    \theta_s(\tau_*^+)\neq 0,\\
    \theta_j(\tau_*^+)\neq \theta_*
    \end{array}$ 
& \emph{Bouncing}       \\ \hline
\end{tabular}
\end{center}
\end{table}

In analogy with other classical  problems in physics (as, among others, the bouncing ball,  the collision of rolling balls, vehicle collision, and rocking motion of rigid block on rigid surfaces), the structural system may lose part of (or even increase in the presence of external forces, as shown later)  its energy during the impact (in the form of heat, noise, dissipation mechanisms, etc). 
Since the difference angles $\theta_s$ and $\theta_j$ are continuous, eqn. (\ref{controtrot}), the potential energy $\Pi$, eqn. (\ref{pot_eng_beta}), is continuous too and therefore the variation in the energy $ 
\mathcal{E}=\mathcal{T}+\Pi$ implies a variation only in the kinetic energy $\mathcal{T}$, eqn. (\ref{kin_eng}).
Following Housner  \cite{Housner1963}, a restitution coefficient $r$ can be introduced as the ratio between the kinetic energies just after and just before the impact,
\begin{equation}\label{rest_def}
	r=\frac{\mathcal{T}(\tau_*^+)}{\mathcal{T}(\tau_*^-)}>0,
\end{equation}
so that the energy difference at the impact can be expressed as
\begin{equation}\label{eng_loss}
	\mathcal{E}(\tau_*^+)-	\mathcal{E}(\tau_*^-)=\mathcal{T}(\tau_*^+)-	\mathcal{T}(\tau_*^-) = -(1-r)\mathcal{T}(\tau_*^-).
	\end{equation}

In some impact problems, inherent assumptions about the motion define the value of the restitution coefficient $r$. For example, by assuming  the pivoting point coincident with the impact point in the rocking motion of a rigid body, a closed form expression for $r$ can be derived through  the law of the conservation of the angular momentum  \cite{Housner1963}. However, the present structural system shows statically indeterminacy at the impact
and therefore a closed-form expression for the restitution coefficient $r$ can not be derived.
An experimental campaign should be performed to effectively overcome the indeterminacy about the restitution coefficient, as well as the  scenario disclosure  after the impact. Since,  experimental measures fall outside the scope of the present research, and these are anyway expected to be  affected by the technical realization of the joints and sliders, and the material composing the structure, the following simplifying assumptions are made 
\begin{enumerate}[label=\roman*.]
    \item \emph{no bouncing} of the impacting layer;
    \item only one layer deformed at every time $\tau$.
\end{enumerate}

Under the two previous assumptions, the  inequality (\ref{theta_conditions})$_3$ constraining the difference angles reduce to
\beq
\theta_1 (\tau) \theta_2(\tau)=0,\qquad \forall\,\, \tau,
\eeq
so that a unique angle measure $\theta(\tau)$ can be introduced to describe in time the two difference angles as
\beq
\left\{
\begin{array}{llll}
\theta_1(\tau)=\theta(\tau),\\[2mm]
\theta_2(\tau)=0,
\end{array}\right.
\qquad \mbox{if}\quad\theta(\tau)\geq 0,\qquad \mbox{and}
\qquad
\left\{
\begin{array}{llll}
\theta_1(\tau)=0,\\[2mm]
\theta_2(\tau)=\theta(\tau),
\end{array}\right.
\qquad \mbox{if}\quad\theta(\tau)\leq 0,
\eeq
and therefore
\begin{center}
    \emph{by excluding bouncing, the present non-smooth two degrees of freedom system\\ can be modeled as  a non-smooth single degree of freedom system.}
\end{center} 

More specifically, the two  equations of motion  \eqref{eomnl} are equivalent to  the following single equation of motion for $\theta(\tau)$

\begin{equation}\label{gov_eqn_theta_long}
	\begin{array}{lcc}
		\left(1+m \right)^{\, \mathsf{H}(\theta (\tau ))}\ddot{\theta}(\tau ) + 
		c ^{\mathsf{H}(\theta (\tau ))}
		\mathcal{C} \; \dot{\theta} (\tau ) +
		k^{\mathsf{H}(\theta (\tau ))} \theta (\tau )  = \\[3mm] 
		 \sin \left(| \theta (\tau )| +\beta _0\right) \left[\left(1+h(\tau)\right)^{\mathsf{H}(\theta (\tau ))} \mathcal{H}(\tau )-\left(1+m\right)^{\mathsf{H}(\theta (\tau ))}\ddot{\zeta}(\tau )\right] \\[3mm] 
		 + \text{sgn}\left[\theta (\tau )\right] \cos \left(| \theta (\tau )| +\beta _0\right) \left[\left(1+v(\tau)\right)^{\mathsf{H}(\theta (\tau ))}\mathcal{V}(\tau ) -\left(1+m\right)^{\mathsf{H}(\theta (\tau ))}\ddot{\nu}(\tau )\right],
	\end{array}
\end{equation}
with the sign, the absolute value, and the Heaviside step $\mathsf{H}$ functions providing the discontinuous character at each impact time $\tau_*$, where $\theta(\tau_*)=0$. The acceleration jump $\salto{0.38}{ \ddot{\theta}(\tau_*)}$ at the impact can be evaluated from the equation of motion (\ref{gov_eqn_theta_long}) as
\begin{equation}
\label{accjump}
\begin{array}{rl}
    \salto{0.38}{ \ddot{\theta}(\tau_*)}= \dfrac{\text{sgn}\left[\theta (\tau_* )\right]}{1+m}  \left\{\right.& \left.(h(\tau_* )-m)\mathcal{H}(\tau_*) \sin\beta_0 + (1-c+m) \,\mathcal{C}\, \salto{0.38}{  \dot\theta(\tau_*)}\right. \\[3mm]
    & \left. + \left[(m+v(\tau_*)+2)\mathcal{V}(\tau_*) - 2 (1+m) \ddot\nu (\tau_*) \right] \cos \beta_0  \right\},
\end{array}
\end{equation}
which shows  that the non-smooth response of the system is not only deriving from  the possible energy (or, equivalently, velocity) change at the impact, and therefore is present even in the case of  kinetic energy (or velocity) continuous at the impact. Moreover, the jump acceleration is not affected by the horizontal ground acceleration $\ddot\zeta$. 

In addition to non-smoothness, in general the system  displays also no symmetry in the  response. However,  a symmetric behaviour in the horizontal load and ground motion can be recovered when the structural system is loaded only with forces at the top layer, ($v(\tau)=h(\tau)=0$), the mass is only present at the top layer ($m=0$), and  the damping and stiffness are  the same for the two layers ($c=k=1$). Indeed, under these assumptions the equation of motion  \eqref{gov_eqn_theta_long} simplifies to\footnote{It is interesting to note that the non-smooth structural system under consideration reduces to an inverted (smooth) pendulum constrained by a linear elastic torsional spring when the undeformed configuration angle is  taken as $\beta_0=\pi/2$,
\begin{equation}\label{gov_eqn_special}
	\begin{array}{lll}
		\ddot{\theta}(\tau ) + \mathcal{C} \; \dot{\theta}(\tau) + \theta (\tau ) = \cos  \theta (\tau ) \left[\mathcal{H}(\tau )-\ddot{\zeta}(\tau )\right] 
		 - \sin \theta (\tau ) \left[\mathcal{V}(\tau ) -\ddot{\nu}(\tau )\right].
	\end{array}
\end{equation}
}
\begin{equation}\label{gov_eqn_theta}
	\begin{array}{lcc}
		\ddot{\theta}(\tau ) + \mathcal{C} \; \dot{\theta} (\tau ) +  \theta (\tau )  =  
		 \sin \left(| \theta (\tau )| +\beta _0\right) \left[\mathcal{H}(\tau ) - \ddot{\zeta}(\tau )\right]  
		 + \text{sgn}\left[\theta (\tau )\right] \cos \left(| \theta (\tau )| +\beta _0\right) \left[\mathcal{V}(\tau ) - \ddot{\nu}(\tau )\right],
	\end{array}
\end{equation}
which implies that the rotation $\theta(\tau)$ has the following   symmetry in the  horizontal load and ground motion (by keeping same  vertical load and ground motion)
\beq
\left.\theta(\tau)\right|_{\mathcal{H}(\tau ), \zeta(\tau ),\mathcal{V}(\tau ), \nu(\tau )}=-\left.\theta(\tau)\right|_{-\mathcal{H}(\tau ), -\zeta(\tau ),\mathcal{V}(\tau ), \nu(\tau )}.
\eeq
Under the considered symmetry assumption, the acceleration jump in  eqn. (\ref{accjump}) reduces into
\begin{equation}
\label{accjump_red}
\begin{array}{ll}
    \salto{0.38}{ \ddot{\theta}(\tau_*)} = \text{sgn}\left[\theta (\tau_* )\right] \left[ \mathcal{C} \, \salto{0.38}{  \dot\theta(\tau_*)}
     + 2 \cos \beta_0 \left[\mathcal{V}(\tau_*) - \ddot\nu (\tau_*) \right] \right],
\end{array}
\end{equation}
confirming that the non-smooth character in the acceleration still persists, although dependent only on the vertical actions, $\mathcal{V}(\tau)$ and $\ddot\nu(\tau)$, and not  not on the horizontal ones, $\mathcal{H}(\tau)$ and $\ddot\zeta(\tau)$.

A further comment on the dissipation at the impact is needed before addressing eqn. \eqref{gov_eqn_theta} in the following Sections  to analyse the non-smooth vibrations of the system under constant and varying loading conditions.
In addition to the restitution factor $r$, the dissipation at the impact can be also expressed through the reduction factor $e\in(0,1]$, which is the multiplier of the velocity just before  the impact, $\dot{\theta}\left(\tau_*^-\right)$, to obtain that just after, $\dot{\theta}\left(\tau_*^+\right)$, namely
\begin{equation}\label{vel_red}
    \dot{\theta}\left(\tau_*^+\right)=e\, \dot{\theta}\left(\tau_*^-\right).
\end{equation}
By considering the kinetic energy $\mathcal{T}$,  eqn. (\ref{kin_eng}), the restitution factor $r$ can be related to the reduction factor $e$ as
\begin{equation}\label{rest_def_vel}
    r= \frac{\left[e\, \dot{\theta} (\tau_*^-) \sin \beta_0 + \dot{\zeta} (\tau_*)\right]^2 + \left[ e\, \dot{\theta}(\tau_*^-) \cos \beta_0 + \dot{\nu} (\tau_*)\right]^2}{\left[ \dot{\theta} (\tau_*^-) \sin \beta_0 + \dot{\zeta} (\tau_*)\right]^2 + \left[ \dot{\theta} (\tau_*^-) \cos \beta_0 + \dot{\nu} (\tau_*)\right]^2},
\end{equation}
showing that the two factors can not be both assumed as constant values, except  when both the ground velocities are null, 
\begin{equation}\label{rest_def_red}
   \dot{\zeta} (\tau_*)=\dot{\nu} (\tau_*)= 0 \qquad \Rightarrow\qquad  r= e^2.
   \end{equation}
  It is evident from eqn. (\ref{rest_def_vel}) that a decrease of kinetic energy $\mathcal{T}$ at the impact defines the two following  equivalent  constraints
  \beq
  r<1 \qquad \Longleftrightarrow \qquad 
    \left[
  (1+e)\dot{\theta}(\tau_*^- )+2\left(\dot{\zeta} (\tau_*)\sin \beta_0 +  \dot{\nu} (\tau_*)\cos \beta_0\right)
  \right]\dot{\theta}(\tau_*^- )>0,
  \eeq
and therefore the condition $e\in(0,1)$ guarantees the energy reduction only in the case of null ground velocities, eqn. (\ref{rest_def_red}). The same phenomenon can be observed also in the classical problem of rocking motion of rigid bodies, which to our best knowledge has not been previously disclosed in literature \cite{Housner1963, Augusti1992, Chopra980, Plaut1996, Makris1999, Makris2001, MASI2019833, Hogan1989}. Indeed, as shown in the next Section, the present system displays some similarities with the rocking motion of rigid bodies in terms of energy balance and behaviour at the impact instant, but  it is fundamentally different for its multistable nature due to presence of restoring forces enabled by the visco-elastic hinges (more details on the energy balance of the rocking rigid body and its similarities with the present system are provided in Appendix \ref{Appen_A}). 
%

In conclusion, the non-smoothness of the present system appears at every impact time ($\tau_*$) and in particular: 
\begin{itemize}
    \item a jump in the acceleration  $\salto{0.38}{ \ddot{\theta}(\tau_*)}$, inherent to the non linearity of the governing equation, is always present;
    \item a jump in the velocity $\salto{0.38}{ \dot{\theta}(\tau_*)}$  may be introduced to represent dissipation mechanisms at impact (as, for example, through the reduction factor $e$).
\end{itemize}

The system is analysed under different loading conditions in the following Sections. First, in order to provide an insight into the influence of the velocity reduction factor, the system has been analysed in Sect. \ref{const_load_vibr}  under constant loading conditions with constant reduction factor less then the unit $e<1$, which as it has been shown in the literature of rocking motion \cite{Dimitrakopoulos2012, Kounadis2013, Vassiliou2015, Bachmann2018, Vassiliou2020}, remains the best approach to introduce the damping mechanism under non-varying loadings. Then,  the analysis is extended in Sect. \ref{forcedvib} to   loading conditions varying in time by modelling  the dissipation mechanisms through viscous damping only (therefore taking a unit reduction factor, $e=1$), because of the  large number of parameters present in the model.

\section{Vibrations at constant vertical loading} \label{const_load_vibr}
Under a constant  vertical load, while all the other forces and the ground motion are null, and for symmetric linear damping
\beq
\mathcal{V}(\tau )=\overline{\mathcal{V}},
\qquad
\mathcal{H}(\tau )=v(\tau)=\ddot{\zeta}(\tau )=\ddot{\nu}(\tau )=0,\qquad
c=1,\eeq
the equation of motion \eqref{gov_eqn_theta_long} reduces to
\begin{equation}\label{gov_eqn_constant_loading}
	\begin{array}{rl}
		\left(1+m \right)^{\, \mathsf{H}(\theta (\tau ))}\ddot{\theta}(\tau ) + 
				\mathcal{C} \; \dot{\theta} (\tau ) +
		k^{\mathsf{H}(\theta (\tau ))} \theta (\tau )  =  \text{sgn}\left[\theta (\tau )\right] \cos \left(| \theta (\tau )| +\beta _0\right)\overline{\mathcal{V}},
	\end{array}
\end{equation}
and the (dimensionless) potential energy $\mathcal{P}(\theta)$ can be evaluated as the elastic energy stored in the deformed spring and the potential of the  load as
\beq
\label{pot_balance0}
\mathcal{P}(\theta)=
k^{\, \mathsf{H}(\theta)}\dfrac{\theta^2}{2} + \overline{\mathcal{V}} \left[\sin\beta_0- \sin \left(\beta _0+|\theta|\right) \right].
\eeq

It is interesting to note that by considering a symmetric system with  null damping ($k=1$, $\mathcal{C}=m=0$), eqn. \eqref{gov_eqn_theta_long} can be rewritten as
\beq
	\ddot{\theta}(\tau ) +\theta(\tau )=  -\text{sgn}\left[\theta (\tau )\right] \,\overline{\mathcal{V}}\, \sin \left(\beta_0-\dfrac{\pi}{2} + | \theta (\tau )| \right), 
\eeq
which differs from the celebrated equation of the free rocking motion for rigid bodies obtained by Housner in 1963 \cite{Housner1963} only for the presence of the linear term in rotation related to the rotational springs (further details are provided in Appendix \ref{Appen_A}).

The dynamics under constant vertical loading and no-bouncing assumption is investigated in this Section through small amplitude vibrations analysis, phase portraits for large amplitude vibrations, and through the realization of a bouncing-like motion from extreme non-symmetric systems.

\subsection{Small amplitude vibrations}\label{small_sec}
Expansion at the first order for small amplitude rotation $\theta(\tau)$ provides the following nonhomogeneous  differential equation with non-constant coefficients
\begin{equation}\label{gov_eqn_constant_loading_lin}
	\begin{array}{rl}
		\left(1+m \right)^{\, \mathsf{H}(\theta (\tau ))}\ddot{\theta}(\tau ) + 
				\mathcal{C} \; \dot{\theta} (\tau ) +
		\left(\overline{\mathcal{V}} \sin \beta _0+k^{\mathsf{H}(\theta (\tau ))}\right) \theta (\tau )  = \text{sgn}\left[\theta (\tau )\right] \overline{\mathcal{V}} \cos \beta_0,
	\end{array}
\end{equation}
which, by considering that  the rotation is continuous in time, consists in a  nonhomogeneous piecewise-linear differential equation with constant coefficients, which is expressed dependently on the rotation sign as
\begin{equation}\label{gov_beta_reduced}
\left\{
\begin{array}{llll}
     \ddot{\theta}(\tau) + 2 \mu_1  \Omega_{0}^{[1]} \, \dot\theta (\tau )+ \left(\Omega_{0}^{[1]}\right)^2 \theta (\tau ) =
     \dfrac{\overline{\mathcal{V}} \cos \beta_0}{1+m},\qquad &\mbox{for} \,\,\theta(\tau)>0,\\[5mm]
     \ddot{\theta}(\tau) + 2 \mu_2  \Omega_{0}^{[2]}  \, \dot\theta (\tau )+ \left(\Omega_{0}^{[2]}\right)^2   \theta (\tau ) =
     -\overline{\mathcal{V}} \cos \beta_0, \qquad &\mbox{for}\,\, \theta(\tau)<0,
\end{array}
\right.
    \end{equation}
where $\Omega_0^{[p]}$ and $\mu_p$ are respectively the \lq apparent' angular frequency and the damping ratio,
\begin{equation}\label{fre_freq_xi}
\begin{array}{llll}
    \Omega_{0}^{[1]} = \sqrt{\dfrac{k+ \overline{\mathcal{V}} \sin \beta_0}{1+m}},
    \qquad
    &\mu_{1} = \dfrac{\mathcal{C}}{2\sqrt{1+m} \sqrt{k+ \overline{\mathcal{V}} \sin \beta_0}},         \\[4mm]
    \Omega_{0}^{[2]}=  \sqrt{1 + \overline{\mathcal{V}} \sin \beta_0},\qquad
     &\mu_{2} =\dfrac{\mathcal{C}}{2\sqrt{1 + \overline{\mathcal{V}} \sin \beta_0}}.
     \end{array}
\end{equation}

Since the derivative of the (dimensionless) potential energy (\ref{pot_balance0}) can be approximated  as
\beq
\label{pot_balanceder}
\dfrac{\mbox{d}\mathcal{P}(\theta)}{\mbox{d} \theta}\approx\left( k^{\, \mathsf{H}(\theta)} + \overline{\mathcal{V}}\sin\beta_0 \right)\theta -\sgn[\theta]\overline{\mathcal{V}}\cos\beta _0.
\eeq
the stability of the undeformed equilibrium configuration is assessed as
\beq\label{stabeunstabletrivial}
\theta=0\,\, \mbox{is }
\left\{
\begin{array}{c}
    \mbox{ stable}\\
\mbox{ unstable}
\end{array}
\right\}\,\,
\mbox{equilibrium if }\,\,
\overline{\mathcal{V}}\cos\beta _0
\left\{\begin{array}{lcl}
<0,\\
>0.
\end{array}\right.
\eeq
In addition to the undeformed equilibrium state  $\theta(\tau)=0$, two adjacent equilibrium states $\theta(\tau)=\theta_A>0$ and  $\theta(\tau)=\theta_B<0$ can be evaluated from eqn. (\ref{gov_beta_reduced}) as
\begin{equation}\label{non-trivial}
    \theta_A=\dfrac{\overline{\mathcal{V}} \cos \beta_0}{k+\overline{\mathcal{V}} \sin \beta_0}>0,\qquad
    \theta_B=-\dfrac{\overline{\mathcal{V}} \cos \beta_0}{1+\overline{\mathcal{V}} \sin \beta_0}<0,
    \end{equation}
which are unstable (or stable) when the undeformed state is stable (or unstable). Indeed, the constraint about  positive  $\theta_A$ and negative $\theta_B$ together with the condition for a stable undeformed state  ($\overline{\mathcal{V}}\cos\beta _0<0$) leads to the following constraint 
\beq\label{minmin}
\min \left\{1,k\right\} + \overline{\mathcal{V}}\sin\beta _0<0
\qquad
\Rightarrow\qquad
\overline{\mathcal{V}}<-\dfrac{\min \left\{1,k\right\}}{\sin\beta_0},
\eeq
which implies a positive  sign in the potential energy differences
\beq\begin{array}{ll}
\mathcal{P}(\theta_A)-\mathcal{P}(0) \approx-\dfrac{\overline{\mathcal{V}}^2\cos^2\beta _0 }{2\left(k+\overline{\mathcal{V}}\sin\beta _0\right)}>0,
\qquad
\mathcal{P}(\theta_B)-\mathcal{P}(0)
\approx
-\dfrac{\overline{\mathcal{V}}^2\cos^2\beta _0 }{2\left(1+\overline{\mathcal{V}}\sin\beta _0\right)}>0,
\end{array}
\eeq
and therefore the instability of the corresponding deformed equilibrium states.

Due to the nature of the small vibration analysis, the two equilibrium configurations $\theta_A$ and $\theta_B$ are relevant only when their modulus is small, condition appearing only if  $\beta_0\approx\pi/2$, for which
\begin{equation}
    \theta_A\approx\dfrac{\overline{\mathcal{V}} }{k+\overline{\mathcal{V}}}\left(\dfrac{\pi}{2}- \beta_0\right),\qquad
    \theta_B\approx-\dfrac{\overline{\mathcal{V}} }{1+\overline{\mathcal{V}}}\left(\dfrac{\pi}{2}- \beta_0\right),
    \end{equation}
under the  constraint (\ref{minmin}), which reduces to
\beq
\overline{\mathcal{V}}<-\min \left\{1,k\right\}.
\eeq
Overall, recalling the restriction $\beta_0\in(0,\pi)$, considering null initial velocity ($\dot\theta(0)=0$) and a non-null position ($\theta(0)=\theta_0\neq 0$), a divergent motion (under small rotations) is realized from the undeformed state $\theta=0$ in the following  cases 
\begin{itemize}
\item when $\beta_0\neq\dfrac{\pi}{2}$ if $
\overline{\mathcal{V}}\left(\dfrac{\pi}{2}-\beta_0\right)>0$;
\item when $\beta_0\rightarrow\left(\dfrac{\pi}{2}\right)^-$
if $
\overline{\mathcal{V}}\left(\dfrac{\pi}{2}-\beta_0\right)<0$ and in addition  if $\overline{\mathcal{V}} < -k$ by taking  $\theta_0>\theta_A$ 
and if $\overline{\mathcal{V}} < -1$ by taking   $\theta_0<\theta_B$;
\item when $\beta_0=\dfrac{\pi}{2}$ if $\overline{\mathcal{V}} < -k$ by taking  $\theta_0>0$ 
and if $\overline{\mathcal{V}} < -1$ by taking   $\theta_0<0$.
\end{itemize}

Recalling the solution from linear oscillators under constant force and with a subcritical damping ($\{\mu_1,\mu_2\}<1$), the vibrations in time  are expressed for initial non-null rotation and null velocity ($\theta_0\neq 0$ and $\dot\theta_0= 0$) by

\begin{equation}\label{linsol_theta1}
    \begin{array}{rr}
        \theta  (\tau) = \mbox{e}^{-\mu_p \, \Omega _0^{[p]}\tau }\left[A_i \cos \left(\sqrt{1-\mu_p^2} \, \Omega _0^{[p]}\tau \right) + B_i \sin \left(\sqrt{1-\mu_p^2} \, \Omega_0^{[p]} \tau \right)\right]
        +(-1)^i \text{sgn}\left[\theta_0\right] \dfrac{\overline{\mathcal{V}} \cos \beta_0}{k^{2-p} +\overline{\mathcal{V}} \sin \beta_0},\\[4mm]
        \mbox{with} \, \,  p=\dfrac{3-(-1)^i \text{sgn}\left[\theta_0\right]}{2},\qquad \mbox{for}
       \, \, \tau\in\left(\tau_*^{[i]},\tau_*^{[i+1]}\right),\, \qquad i\in \mathbb{N}_0,  
    \end{array}
\end{equation}
where $\tau_*^{[0]}=0$, $\tau_*^{[i]}$ represents the time when the $i$-th (non-bouncing) impact occurs ($i\in\mathbb{N}$),  while $A_i$ and $B_i$ are  integration constants defining the motion before the $(i+1)$-th impact and after the previous one to be  evaluated as iterative procedure starting  from the initial conditions, $\theta  (0) = \theta_0$ and $\dot{\theta}(0)=0$, for $i=0$ and following the   condition of continuous rotation  and discontinuous velocity at the $i$-th impact for $i\geq 1$, given by
\begin{equation}
    \theta \left(\tau_*^{[i]-}\right) = \theta \left(\tau_*^{[i]+}\right)  = 0,\qquad
    \dot{\theta} \left(\tau_*^{[i]+}\right)  = e \, \dot{\theta}  \left(\tau_*^{[i]-}\right) ,\qquad \,\,i\in \mathbb{N}.
\end{equation}

Restricting the attention to undamped symmetric systems ($m=\mathcal{C}=0$, $k=1$), for which $\Omega_{0}^{[p]} = \Omega_{0}$ and $\mu_{p}=0$ ($p=1,2$), the small vibrations at constant vertical loading reduce to
\begin{equation}\label{linsol_theta2}
    \begin{array}{rr}
        \theta  (\tau) = A_i \cos \left( \Omega _0 \tau \right) + B_i \sin \left(\Omega_0 \tau \right)
        +(-1)^i \text{sgn}\left[\theta_0\right] \dfrac{\overline{\mathcal{V}} \cos \beta_0}{1+\overline{\mathcal{V}} \sin \beta_0},\qquad \mbox{for}
       \, \, \tau\in\left(\tau_*^{[i]},\tau_*^{[i+1]}\right),\, \qquad i\in \mathbb{N}_0,
    \end{array}
\end{equation}
with the two non-trivial equilibrium rotations (\ref{non-trivial}) becoming 
\begin{equation}
    \theta_A=-\theta_B=\dfrac{\overline{\mathcal{V}} \cos \beta_0}{1+\overline{\mathcal{V}} \sin \beta_0},
\end{equation}
which are relevant only when $\beta_0\approx\pi/2$ and are unstable when 
\begin{equation}\label{condunst}
\overline{\mathcal{V}}<-1.
\end{equation}
By assuming that the  dissipation does not appear even at the impact ($e=1$), the motion becomes  periodic  with a dimensionless period  $\Te_{0}$, different from $2\pi /\Omega_0$ whenever $\overline{\mathcal{V}}\neq 0$,  given by
\begin{equation}\label{t-beta}
\begin{array}{ll}
    \Te_{0}(\theta_0) = \dfrac{4}{\sqrt{1 + \overline{\mathcal{V}} \sin \beta_0}}\arccos\left(\dfrac{1}{1-  \dfrac{1+\overline{\mathcal{V}} \sin \beta_0}{\overline{\mathcal{V}}\cos \beta _0}|\theta _0|}\right),
    \end{array}
\end{equation}
which, from its expansion  at small rotation ($\theta_0\approx0$),  under the  constraint $|\theta_0|\ll \left| \overline{\mathcal{V}}\cos\beta_0/(1+\overline{\mathcal{V}}\sin\beta_0)\right|$, for $\beta_0\not\approx\pi/2$ results proportional to the square root of the initial perturbation $\theta_0$, similarly to the dynamics of a rocking rigid body \cite{Housner1963} (see Appendix \ref{Appen_A})
\begin{equation}\label{t-beta_exp}
    \begin{array}{ll}
    \Te_{0}(\theta_0) \approx  \dfrac{4\sqrt{2|\theta _0|}}{\sqrt{-\overline{\mathcal{V}} \cos \beta _0}},
    \end{array}
\end{equation}
confirming the inequality (\ref{stabeunstabletrivial}) for having  no divergent motion from the undeformed state $\theta=0$.
Moreover, in the case when  $\beta_0\approx\pi/2$, the non-trivial equilibrium configuration can be approximated by  $\theta_A\approx\left(\pi/2-\beta_0
    \right)\overline{\mathcal{V}}
/(1+\overline{\mathcal{V}})$ and the following expansion for the dimensionless period  $\Te_{0}$ can be obtained
\begin{equation}\label{t-beta_exp2}
    \begin{array}{ll}
    \Te_{0}(\theta_0) \approx\dfrac{4}
    {\sqrt{1+\overline{\mathcal{V}}} } \arccos\left[
           \dfrac{
    1
    }{    
1 -  \dfrac{
1+\overline{\mathcal{V}}}{
\overline{\mathcal{V}}}
\dfrac{|\theta_0|}{\frac{\pi}{2}-\beta_0}
}
           \right],
    \end{array}
\end{equation}
further confirming  that the divergent motion (under the small rotation assumption) from the undeformed state ($\theta=0$) is realized under the conditions obtained from the potential energy analysis discussed before.

Extending now the small amplitude vibrations to the presence of impact dissipation ($e \leq 1$), the oscillations lose periodicity by decreasing amplitude and period after every impact.
Indeed, through  energy balance, the (small) oscillation amplitude $\theta_{D}^{\{n\}}=\theta\left(\tau_*^{[2(n-1)]}\right)$ at the  beginning of the $n$-th oscillation (occurring from  $\tau_*^{[2(n-1)]}$ and $\tau_*^{[2(n)]}=\tau_*^{[2(n-1)]}+\Te_{D}^{\{n\}}$, therefore  encompassing two impacts, $n\in\mathbb{N}$) 
can be approximated as
 \beq\label{approx1}
 \theta_{D}^{\{n\}}\approx e^{4n}\theta_0,
 \eeq
 while the $n$-th cycle  period $\Te_{D}^{\{n\}}$ can be
 approximated by\footnote{Since the oscillation period decreases as effect of each impact, the time interval between two impacts decreases too, namely
 $$
 \tau_*^{[2n]}-\tau_*^{[2n-1]}< \tau_*^{[2n-1]}-\tau_*^{[2(n-1)]},\qquad n\in\mathbb{N}.
 $$}
 \begin{equation}\label{reduced_period}
    \begin{array}{cc}
        \Te_{D}^{\{n\}} \approx e^{2(n-1)}  \Te_{D}^{\{1\}},\qquad\mbox{with}\qquad\Te_{D}^{\{1\}}\approx \dfrac{(1+e)^2}{4}\Te_0
        \leq \Te_0, 
    \end{array}
\end{equation}
showing the decrease of both the amplitude $ \theta_{D}^{\{n\}}$ and  period $\Te_{D}^{\{n\}} $ with the increasing number of impact cycles, similarly to the rocking motion of rigid bodies. More specifically, it is noted that the period $\Te_D^{\{n\}}$ approaches zero in the limit of infinite number $n$ of oscillations and, interestingly, the oscillatory motion ends at a finite time $\tau_{\mbox{\scriptsize{fin}}}$, because the corresponding  time $\tau_*^{[2n]}$ for infinite $n$ is given by the following convergent geometric series (for $e<1$)
\beq\label{finaltime}
\tau_{\mbox{\scriptsize{fin}}}=\lim_{n\rightarrow\infty}\tau_*^{[2n]}=\sum_{p=1}^\infty \Te_{D}^{\{p\}}\approx\dfrac{1+e}{4(1-e)}\Te_0.
\eeq

\subsection{Phase portraits of  vibrations at large amplitude}\label{phase_sec}

The vibrations for a system under a constant vertical loading are here  addressed without any restriction about small rotations. The analysis is referred to a system with $\beta_0=80^\circ$ and loaded only with a constant vertical load  $\overline{\mathcal{V}}=-1.5/\sin\beta_0\approx -1.523$, realizing  multistability through three\footnote{It is noted that the tetrastability  presented in \cite{hima22} reduces here to tristability because of the assumption about  no more than one deformed layer at any time.}  stable equilibrium states ($\theta=0$, $\theta=\theta_I>0$ and $\theta=\theta_{II}<0$), while for simplicity a null mass ratio is assumed ($m=0$). Considering specific initial rotation and velocity, the numerical integration of the equation of motion \eqref{gov_eqn_constant_loading} can be performed and  the trajectories within the phase plane $\theta-\dot\theta$ defined. These trajectories show how the motion evolves in time and in particular allows for the definition of \emph{separatrices}, namely
lines separating conditions in regions corresponding to  qualitatively different motion. Moreover, all the trajectories display a  discontinuity in their tangent when crossing the switching condition $\theta=0$.

\begin{figure}[!ht]
\centering
\includegraphics[width=\textwidth]{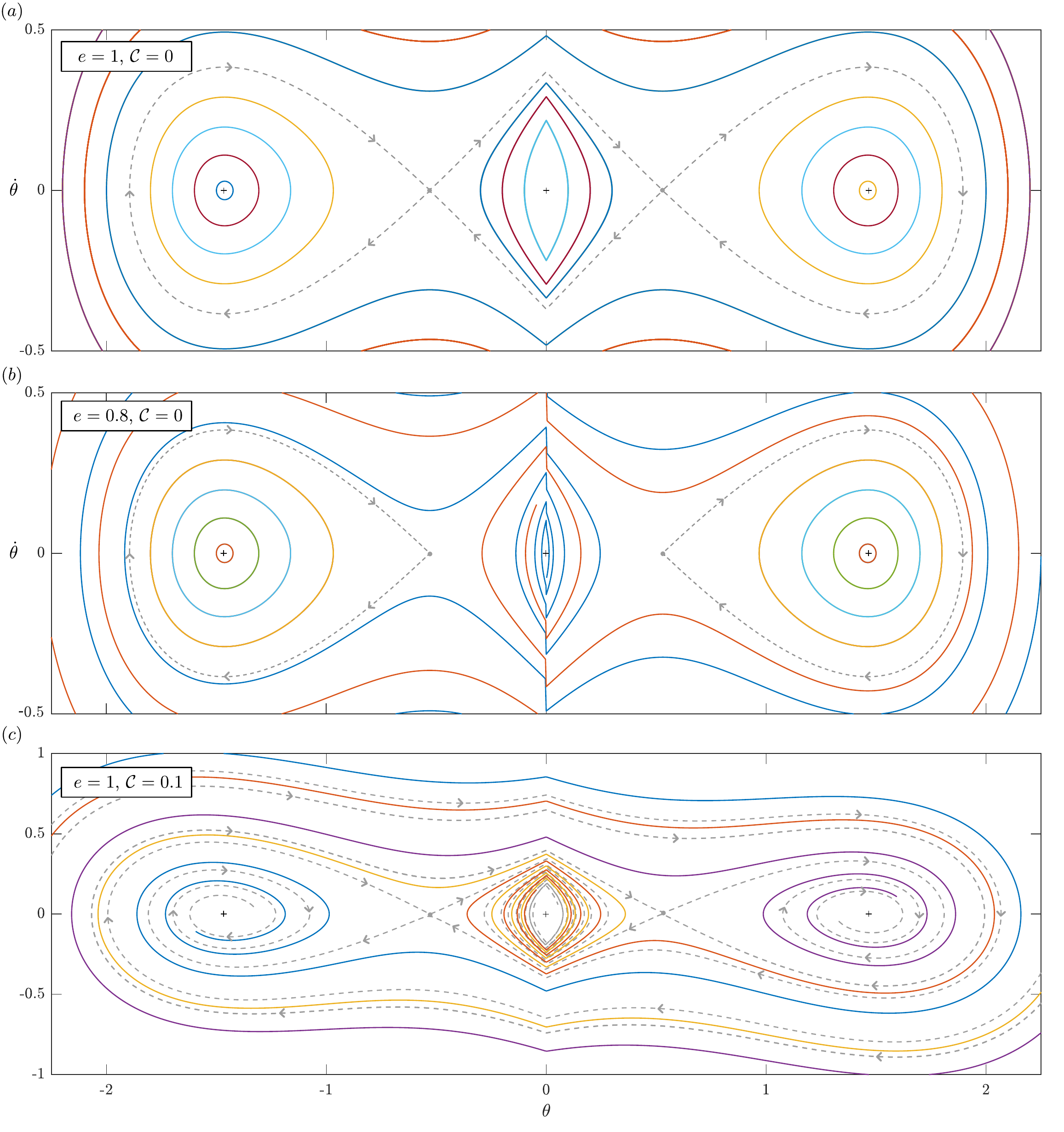}
\caption{Phase portrait $\theta(\tau)-\dot{\theta}(\tau)$ for the symmetric system ($k=1$, $m=0$) under a  constant vertical load $\overline{\mathcal{V}}=-1.523$ and for $\beta_0=80^\circ$ with varying dissipation source: $(a)$ no dissipation ($e=1$, $\mathcal{C}=0$), $(b)$ dissipation  through impact only ($e=0.8$, $\mathcal{C}=0$), and $(c)$ dissipation  through viscous damping only  ($e=1$, $\mathcal{C}=0.1$).}
\label{phase-diagram-sym}
\end{figure}

\begin{figure}[!ht]
\centering
\includegraphics[width=\textwidth]{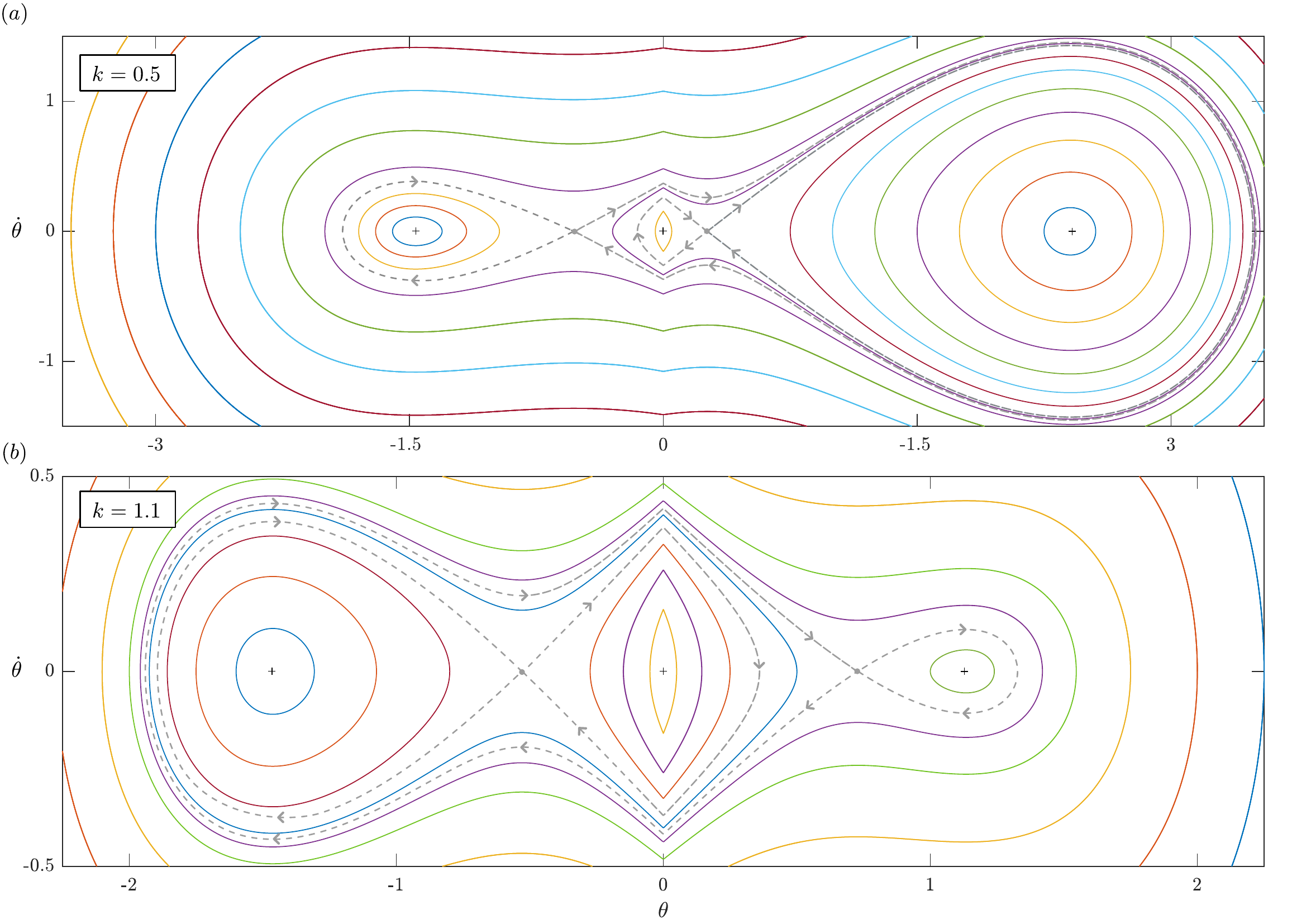}
\caption{
Phase portrait $\theta(\tau)-\dot{\theta}(\tau)$ for  non-symmetric conservative systems ($e=1$, $m=\mathcal{C}=0$) under a  constant vertical load $\overline{\mathcal{V}}=-1.523$ and for $\beta_0=80^\circ$ with varying stiffness ratio: $(a)$ $k=0.5$ and $(b)$  $k=1.1$.}
\label{phase-diagram-asym}
\end{figure}

The phase portraits for  a symmetric system ($k$=1, for which $\theta_I=-\theta_{II}=1.462$) are reported in Fig. \ref{phase-diagram-sym}  with varying the dissipation source, more specifically: 
\begin{itemize}
\item[ ]
    \begin{itemize}
        \item[ Fig. \ref{phase-diagram-sym} $a$)] no dissipation  ($\mathcal{C}=0$, $e=1$).  The system is conservative and the phase portrait is symmetric with respect to both the $\theta$ and $\dot\theta$ axes. The separatrices (reported as dashed gray line) define three bounded regions, each one containing one attractor (equilibrium configuration) only. Therefore, the periodic oscillating motion is around only one attractor if the initial condition is within the corresponding separatrix curve, otherwise around all of the three attractors if the condition is outside;
        \item[ Fig. \ref{phase-diagram-sym} $b$)] dissipation  through impact only ($\mathcal{C}=0$, $e=0.8$). With respect to the previous conservative case, the trajectories are modified only when crossing the switching condition ($\theta=0$), where a jump discontinuity appears. It follows that the portion of separatrix line  bounding the region encompassing the attractor at the origin ($\theta=0$)  disappears and the related region  merges with the unbounded one to generating a decaying motion towards the undeformed configuration, while the other two bounded regions still define periodic motion around the respective non-trivial attractor ($\theta=\theta_I$ and $\theta=\theta_{II}$) and remain unaffected;
        \item [ Fig. \ref{phase-diagram-sym} $c$)] dissipation  through  viscous damping only ($\mathcal{C}=0.1$, $e=1$). By considering that energy is  continuously dissipated in time, every trajectory defines a decaying (non-periodic) oscillatory motion towards one of the three attractors. Therefore, differently from the two previous cases, no trajectory is bounded and, although approaching a different attractor at infinite time, each trajectory describes  oscillations around all of the three attractors for some time interval.
    \end{itemize}
\end{itemize}    
For completeness, the  response for a conservative non-symmetric  system ($\mathcal{C}=0$, $e=1$) is shown in Fig. \ref{phase-diagram-asym} for two different values of stiffness ratio $k$. 

With respect to the symmetric ($k=1$) case,  only the portion of trajectories within the half-plane $\theta>0$, and correspondingly the positive equilibrium configuration $\theta_I$, is modified. 
In particular:
\begin{itemize}
\item[ ]
    \begin{itemize}
        \item [Fig. \ref{phase-diagram-asym} $a)$] lower layer softer than the upper one  ($k=0.5$). The positive value of non-trivial equilibrium increases ($\theta_I=2.405$) as the size of the region  defined by the separatrix encompassing it;
        \item [Fig. \ref{phase-diagram-asym} $b)$] lower layer stiffer than the upper one ($k=1.1$). The positive value of non-trivial equilibrium decreases ($\theta_I=1.384$) and, as  a consequence, the corresponding separatrix region shrinkes.
    \end{itemize}
\end{itemize}
Moreover, differently from the conservative symmetric case ($k=e=1$, $\mathcal{C}=0$), the separatrices in  the conservative non-symmetric system show the possibility of periodic oscillations around  two (instead of only one or three) equilibrium configurations, corresponding to  $\theta_I$ when $k<1$ ($\theta_{II}$ when $k>1$) and $\theta=0$. 

\subsection{Bouncing-like motion from extreme non-symmetric systems under no-bouncing assumption}\label{bounc_sec}
Despite the no-bouncing assumption at impact, non-symmetric systems may display a bouncing-like motion when characterized by  extreme stiffness ratio. For the sake of simplicity, this behaviour is shown by analyzing conservative systems, therefore in absence of any dissipative mechanism ($e=1$, $\mathcal{C}=0$) and loadings varying in time, although these are  not expected to qualitatively modify the   response. It follows that the  total energy is conserved and the motion is periodic, therefore defining the  maximum and minimum rotations as 
\beq
\theta_{\min}=\min_{\tau} \{\theta(\tau)\}<0,
\qquad
\theta_{\max}=\max_{\tau} \{\theta(\tau)\}>0,
\eeq 
where velocity annihilates, the total energy matching at these two states reduces to the following (dimensionless) potential energy matching
\beq
\label{pot_balance}
k\dfrac{\theta_{\min}^2}{2} -\overline{\mathcal{V}} \sin \left(\beta _0+\theta_{\min}\right)= 
\dfrac{\theta_{\max}^2}{2}-\overline{\mathcal{V}} \sin \left(\beta _0-\theta_{\max}\right),
\eeq
establishing a nonlinear relation  between the maximum and minimum rotation values, $\theta_{\max}$ and $\theta_{\min}$.
Introducing the  parameter $P_R$ representing the peak ratio and defined as
\beq
P_R\left(\overline{\mathcal{V}},k\right)=
\min\left\{
\dfrac{\theta_{\max}}{\left|\theta_{\min}\right|},
\dfrac{\left|\theta_{\min}\right|}{\theta_{\max}}
\right\}\in[0,1],
\eeq
a bouncing-like motion is realized in the limit of vanishing $P_R$. 

It is noted that, from the energy balance \eqref{pot_balance},  the peak ratio $P_R$ satisfies the following identity
\beq\label{Prprop}
\left.P_R\left(\overline{\mathcal{V}},k\right)\right|_{\theta_{\min}=-|\theta_0|} =  \left.
P_R\left(\dfrac{\overline{\mathcal{V}}}{k},\dfrac{1}{k}\right)\right|_{\theta_{\max}=|\theta_0|},\qquad k\geq 1.
\eeq

\paragraph{Null vertical load $\overline{\mathcal{V}}=0$.} 
When the vertical load is absent, the equation of motion \eqref{gov_eqn_constant_loading} becomes homogeneous,
\begin{equation}\label{gov_eqn_constant_loading22}
	\begin{array}{rl}
		\left(1+m \right)^{\, \mathsf{H}(\theta (\tau ))}\ddot{\theta}(\tau ) + 
		k^{\mathsf{H}(\theta (\tau ))} \theta (\tau )  =  0,
	\end{array}
\end{equation}
and the energy matching (\ref{pot_balance}) leads to 
\beq
\theta_{\max}=-\sqrt{k} \theta_{\min},
\eeq
from which the peak ratio $P_R$ can be analytically evaluated as
\begin{equation}\label{quasi-bounce-peak}
	P_R(k) = 
	\min\left\{
\dfrac{1}{\sqrt{k}},
\sqrt{k}
\right\}.
\end{equation}
Likewise, the period of oscillation $\Te_0$, in terms of the dimensionless time $\tau=t\sqrt{M_2/K_2}$, is given by the sum of the time duration $\Te_0^{[1]}$ of the oscillation for positive rotation and by  $\Te_0^{[2]}$ for the negative ones, respectively given  by
\begin{equation}\label{period_case1}
	\begin{array}{cc}
		\Te_0^{[1]}(k,m) =  \pi \sqrt{\dfrac{m+1}{k}} ,\qquad
		\Te_0^{[2]} =  \pi,
	\end{array}
\end{equation}
and therefore the period of oscillation $\Te_0$ follows as  
\begin{equation}\label{period_case2}
	\begin{array}{cc}
		\Te_0(k,m) =  \pi\left(1 + \sqrt{\dfrac{m+1}{k}} \right).
	\end{array}
\end{equation}
It can be observed from eqn. (\ref{quasi-bounce-peak}) that  the peak ratio $P_R$ vanishes when the stiffness ratio $k$ takes extremely small or large values,
\beq
\lim_{k\rightarrow 0}
P_R(k)=
\lim_{k\rightarrow\infty}
P_R(k)=0,\eeq
resembling  a bouncing-like motion. 
\begin{figure}[!ht]
\centering
\includegraphics[width=\textwidth]{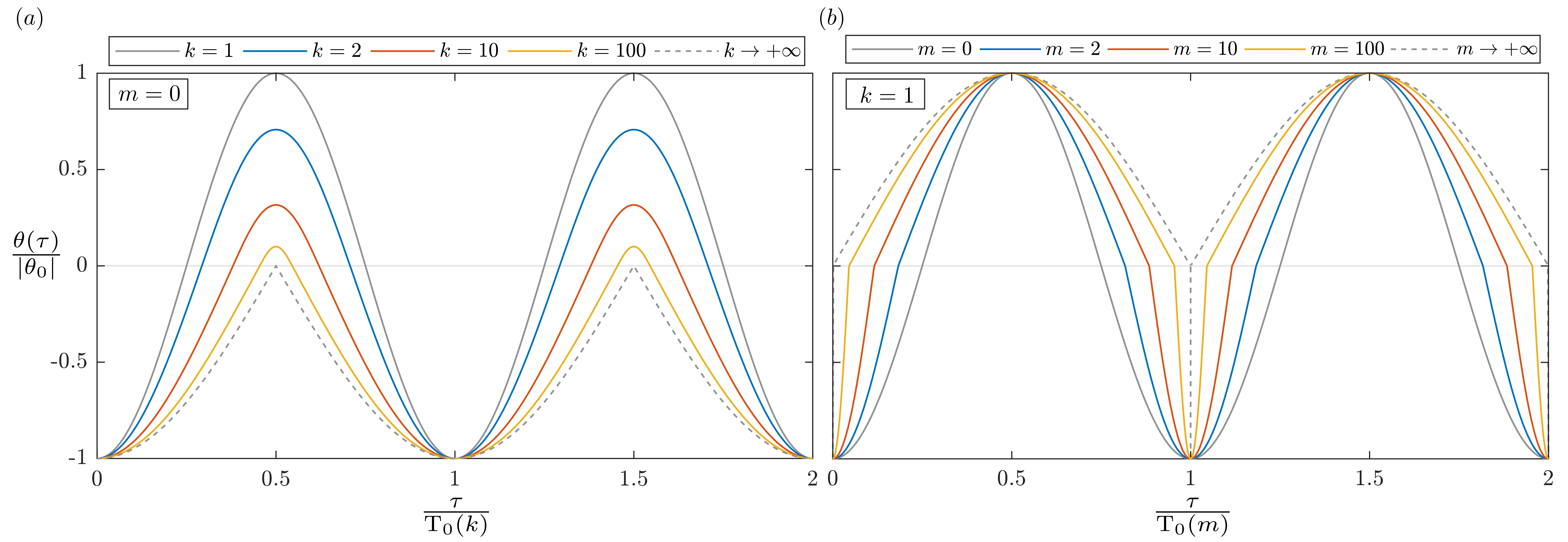}
\caption{Time histories $\theta-\tau/\Te_0(k,m)$ of  a conservative system ($\mathcal{C}=0$, $e=1$) with $\beta_0=80^\circ$, under null  vertical  and horizontal  loads, and with ($a$) stiffness ratio $k\geq 1$ and null mass ratio  $m=0$ and ($b$) unit stiffness ratio $ k=1$ and non-null mass ratio $m\geq 0$.}
\label{quasi bouncing}
\end{figure}

It is interesting to observe that the corresponding periods of oscillation $\Te_0$ are coincident with the  oscillation time corresponding to only one (among the two) sign of rotation, 
\beq\begin{array}{lll}
\displaystyle\lim_{k\rightarrow 0}
\dfrac{\Te_0^{[2]}}{\Te_0^{[1]}(k,m)}=0\qquad\Rightarrow \qquad
\lim_{k\rightarrow 0}\Te_0=\lim_{k\rightarrow 0}\Te_0^{[1](k,m)}\rightarrow\infty,\\[6mm]
\displaystyle\lim_{k\rightarrow\infty}
\dfrac{\Te_0^{[1]}(k,m)}{\Te_0^{[2]}}=0
\qquad\Rightarrow \qquad
\lim_{k\rightarrow \infty}\Te_0=\Te_0^{[2]}=\pi.
\end{array}
\eeq
It is worth to mention that the  infinite  period of oscillation $\Te_0$ at vanishing stiffness ratio $k$ appears only because of the adopted  time normalization (namely, with respect to $K_2$ and $M_2$ instead of $K_1$ and $M_1$).
It also interesting to note that, although $m$ does not affect the peak ratio $P_R$, extremely large values of $m$  define vanishing duration of oscillations for negative sign of $\theta(\tau)$, namely
\beq
\lim_{m\rightarrow\infty}
\dfrac{\Te_0^{[2]}}{\Te_0^{[1]}(k,m)}=0
\qquad\Rightarrow \qquad
\lim_{m\rightarrow\infty}\Te_0=\lim_{m\rightarrow\infty}\Te_0^{[1](k,m)}\rightarrow\infty.
\eeq
Therefore, the case of large values of $m$ (at finite values of $k$) does not represent a real bouncing-like motion, although the oscillation phase involving negative rotation has vanishing duration and therefore it can be interpreted as an impulse to the system in the proper timescale.
By considering only one source of non-symmetry in terms of stiffness ($k\geq 1$, $m=0$) or mass ($k=1$, $m \geq 0$) ratio, the rotation as a function of the dimensionless time $\tau/\Te_0(k,m)$ is reported in Fig. \ref{quasi bouncing} for $\beta_0=80^\circ$.
The case $k<1$ is not reported since it is essentially the mirrored version of Fig. \ref{quasi bouncing}-$a$) about the (logarithmic) horizontal axis of the case with the corresponding reciprocal stiffness $1/k>1$, as described by  \eqref{Prprop}. 
\begin{figure}[!ht]
\centering
\includegraphics[width=\textwidth]{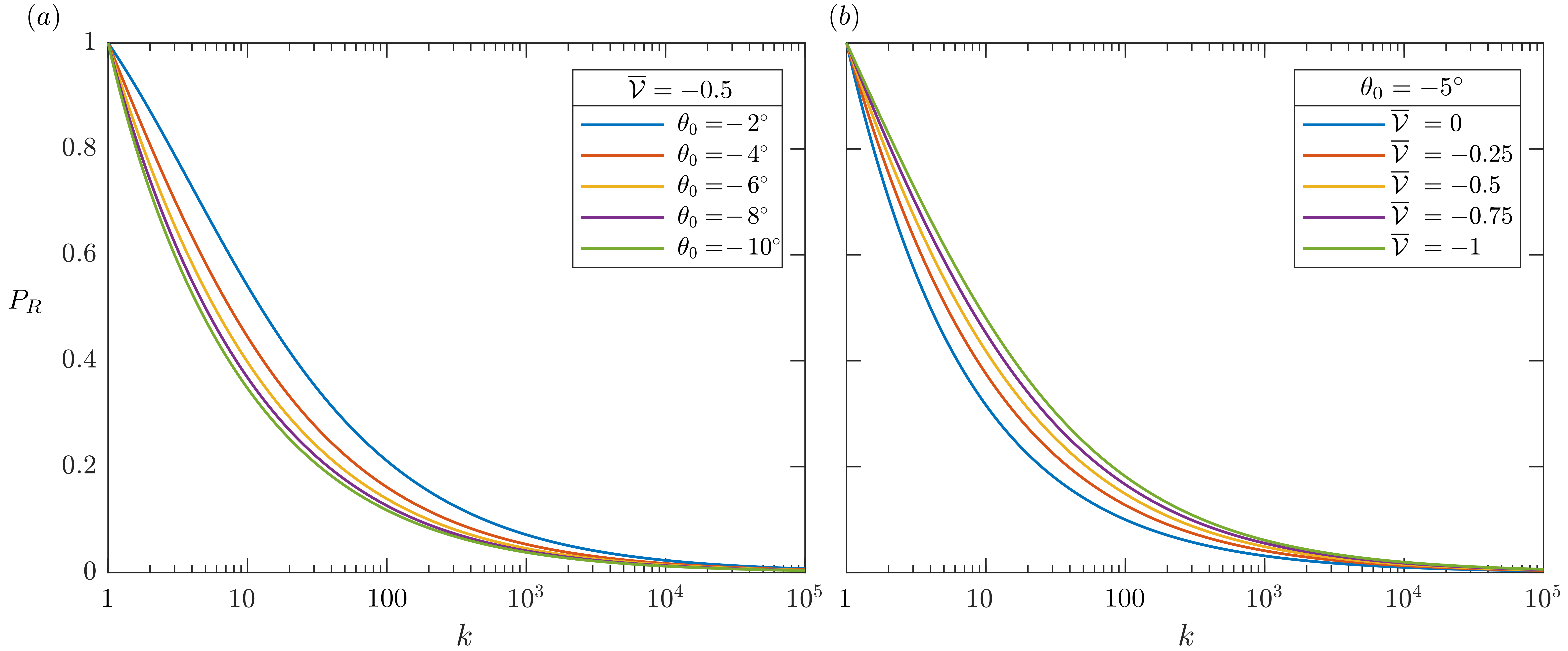}
\caption{Peak ratio $P_R$ for a conservative  system  ($\mathcal{C}=0$, $e=1$) with $\beta_0=80^\circ$, $m=v=c=0$ at varying the stiffness ratio $k\geq 1$ for $(a)$ a vertical load $\overline{\mathcal{V}}=-0.5$ and different values of the initial rotation $\theta_0=-\{2,4,6,8,10\}^\circ$ and $(b)$ an initial rotation $\theta_0=-5^\circ$ and different constant vertical loads $\overline{\mathcal{V}}=-\{0,0.25,0.5,0.75,1\}$.}
\label{bouncing-like_response}
\end{figure}
\paragraph{Non-null constant vertical load $\overline{\mathcal{V}}<0$.} 
The realization of the bouncing-like motion in the case of extreme stiffness ratio values $k$ is here shown to occur also in  the presence of a negative vertical load $(\overline{\mathcal{V}}<0)$.
In this case, the peak ratio $P_R$ can be  obtained only numerically by solving the nonlinear  energy matching \eqref{pot_balance}. The peak ratio $P_R$  for a conservative system ($\mathcal{C}=0$, $e=1$, $\beta_0=80^\circ$) is reported in Fig. \ref{bouncing-like_response} with varying  the stiffness ratio $k$ for $(a)$ constant  vertical load  $\overline{\mathcal{V}}=-0.5$ and
different initial rotation $\theta_0<0$ and for $(b)$ constant initial rotation $\theta_0=-5^\circ$  and different vertical loads $\overline{\mathcal{V}}$.
In both cases, the monotonic decrease of $P_R$ is observed at increasing values of the stiffness ratio $k\geq 1$. Therefore the peak ratio approaches null values, representing  a bouncing-like motion, in the limit of  infinitely large stiffness ratio, $P_R\left(\overline{\mathcal{V}},k\rightarrow\infty\right)= 0$. Due to the symmetry property \eqref{Prprop}, a bouncing-like motion is also realized in the limit of vanishing stiffness ratio, $\lim_{k\rightarrow 0}P_R\left( k\overline{ \mathcal{V}}, k \right)= 0$. 

\section{Vibrations under loading varying in time}\label{forcedvib}
The vibrations of the symmetric system ($m=v(\tau)=h(\tau)=0$, $k=c=1$) dissipated  through viscous damping only  ($\mathcal{C}>0$, $e=1$) are analysed under different harmonic loads. 
\begin{figure}[!ht]
\centering
\includegraphics[width=\textwidth]{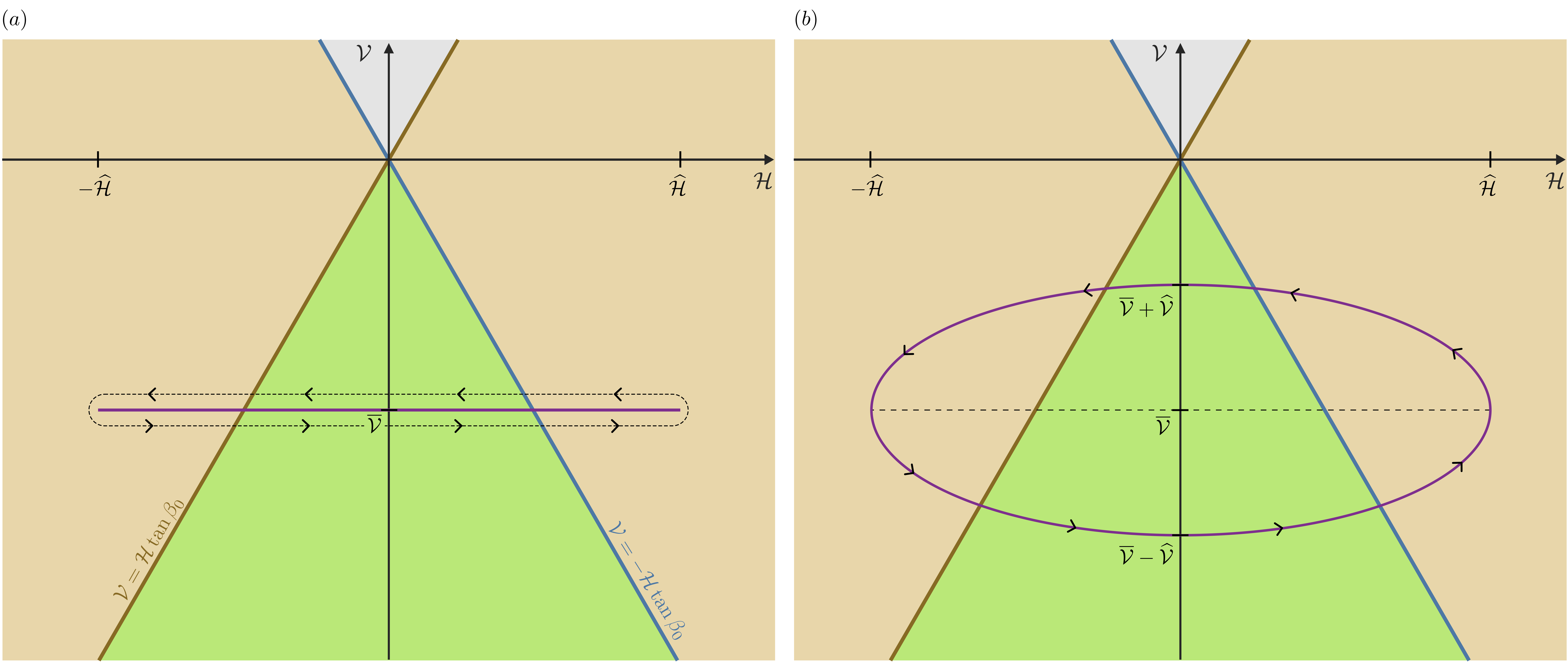}\caption{Loading paths in time within the  $\mathcal{H}$-$\mathcal{V}$ plane for $(a)$  harmonic horizontal force  at constant vertical load $\overline{\mathcal{V}}$, eqn. \eqref{loadpatha}, and $(b)$  harmonic horizontal  and vertical  loads,  eqn. \eqref{loadpathb}. By modifying the loading amplitudes, the loading paths are considered within or intersecting the  quasi-static monostable domain associated with  the undeformed configuration (and reported in the sketch for $\beta_0\in(0,\pi/2)$),  defined by the loading pairs satisfying  $\mathcal{V}\cos\beta_0<-|\mathcal{H}|\sin\beta_0$. }
\label{fig:loading_paths}
\end{figure}
In particular, the results are obtained through the  numerical integration of the eqn. \eqref{gov_eqn_theta} under the following loading conditions:
\begin{enumerate}[label=\roman*.]
    \item Harmonic horizontal  load at constant vertical load (Fig. \ref{fig:loading_paths}-$a$)
    \begin{equation}\label{loadpatha}
       \mathcal{V}(\tau)=\overline{\mathcal{V}},\qquad \mathcal{H}(\tau)=\widehat{\mathcal{H}} \sin (\Omega \tau),
    \end{equation}
    \item  Harmonic horizontal  ground motion at constant vertical load
        \begin{equation}\label{loadpathaa}
       \mathcal{V}(\tau)=\overline{\mathcal{V}},\qquad \mathcal{\zeta}(\tau)=\widehat{\mathcal{\zeta}} \sin (\Omega \tau),
    \end{equation}
    which, from eqn. \eqref{gov_eqn_theta}, is equivalent to loading condition (i.) by considering $\widehat{\mathcal{H}}=-\Omega^2\widehat{\mathcal{\zeta}}$,
    \item Harmonic horizontal and vertical force at the second layer (Fig. \ref{fig:loading_paths}-$b$)
    \begin{equation}\label{loadpathb}
       \mathcal{H}(\tau)=\widehat{\mathcal{H}} \sin (\Omega \tau)
       \qquad
       \mathcal{V}(\tau)=\overline{\mathcal{V}}+\widehat{\mathcal{V}} \cos (\Omega \tau).
        \end{equation}
\end{enumerate}

Unless otherwise specified throughout the Section, the constant vertical load applied on the second layer is assumed as $\overline{\mathcal{V}}=-0.5$ the initial configuration angle  as $\beta_0=80^\circ$ and the linear viscous damping as $\mathcal{C}=0.1$.  The remaining  parameters of loading conditions are specified in the relevant parts of the text.

The dynamic behaviour of the unit structural cell is investigated in terms of frequency response curves and bifurcation diagrams,  obtained from the numerical integration of eqn. \eqref{gov_eqn_theta} through the commercial code  MATLAB by exploiting the continuation toolbox MATCONT \cite{matcont} (employing $\texttt{ode45}$ solver, $\left\{\texttt{Relative Tolerance}\right.$,  $\left. \texttt{Absolute Tolerance},\texttt{Initial Time Step}\right\}\approx10^{-8}$ and $\texttt{Maximum Time Step}$ $\approx10^{-4}$). Considering the inherent non-smoothness of the equation of motion \eqref{gov_eqn_theta} and the prerequisites for using the continuation toolbox, a necessary adjustment is introduced to smoothen  the sign function $\mbox{sgn}[\theta(\tau)]$ by approximating it through the following exponential function 
$q(\theta(\tau),\mathcal{B})$ 
\begin{equation}\label{smoothing}
    q(\theta(\tau),\mathcal{B})=  \dfrac{e^{\mathcal{B} \theta(\tau)}-1}{e^{\mathcal{B} \theta(\tau)}+1},
\end{equation}
by considering large values of the dimensionless quantity $\mathcal{B}$,  selected to not compromise  the accuracy of the limit cycle  while the continuation method converges. To verify the accuracy of the model and selected parameter $\mathcal{B}$, the parts corresponding to stable motions of the MATCONT outputs  have been independently confirmed through a direct integration of the non-smooth equation of motion in MATLAB (without any smoothing of the $\sgn$ function) by means of the same  solver and  options   adopted in the continuation method.

By introducing $\theta_{\rm{st}}$ as the difference angle corresponding to the quasi-static equilibrium under the loads  $\mathcal{H}=\widehat{\mathcal{H}}$ and $\mathcal{V}=\overline{\mathcal{V}}$ when the system is harmonically loaded and $\theta_{\rm{eq}}$ as the  equivalent rotation angle when the system is under an harmonic  ground motion with amplitude $\widehat{\zeta}$ 
\begin{equation}
    \theta_{\rm{eq}}=\arccos\left(\cos\beta_0-\widehat{\mathcal{\zeta}}\right)-\beta_0,
\end{equation}
the dynamic response is reported in terms of dimensionless rotation $\rm{max}_\tau|\theta(\tau)|/\theta_{\rm{st}}$ or $\rm{max}_\tau|\theta(\tau)|/\theta_{\rm{eq}}$.
As graphical convention, the sets of stable configuration are drawn with a continuous blue line while the unstable ones  with a dashed red line; frequency range displaying quasi-periodic (QP) and    chaotic (CH) responses are  highlighted through  light blue and light orange bands, respectively. Moreover, in  some of the reported  spectra, a horizontal dashed green line appears at the difference angle $\overline{\theta}^{[1]}=\pi-\beta_0=110^\circ$, corresponding to the smallest rotation for which the layer of parallelogram shape reduces into a straight line.

\begin{figure}[!ht]
\centering
\includegraphics[width=\textwidth]{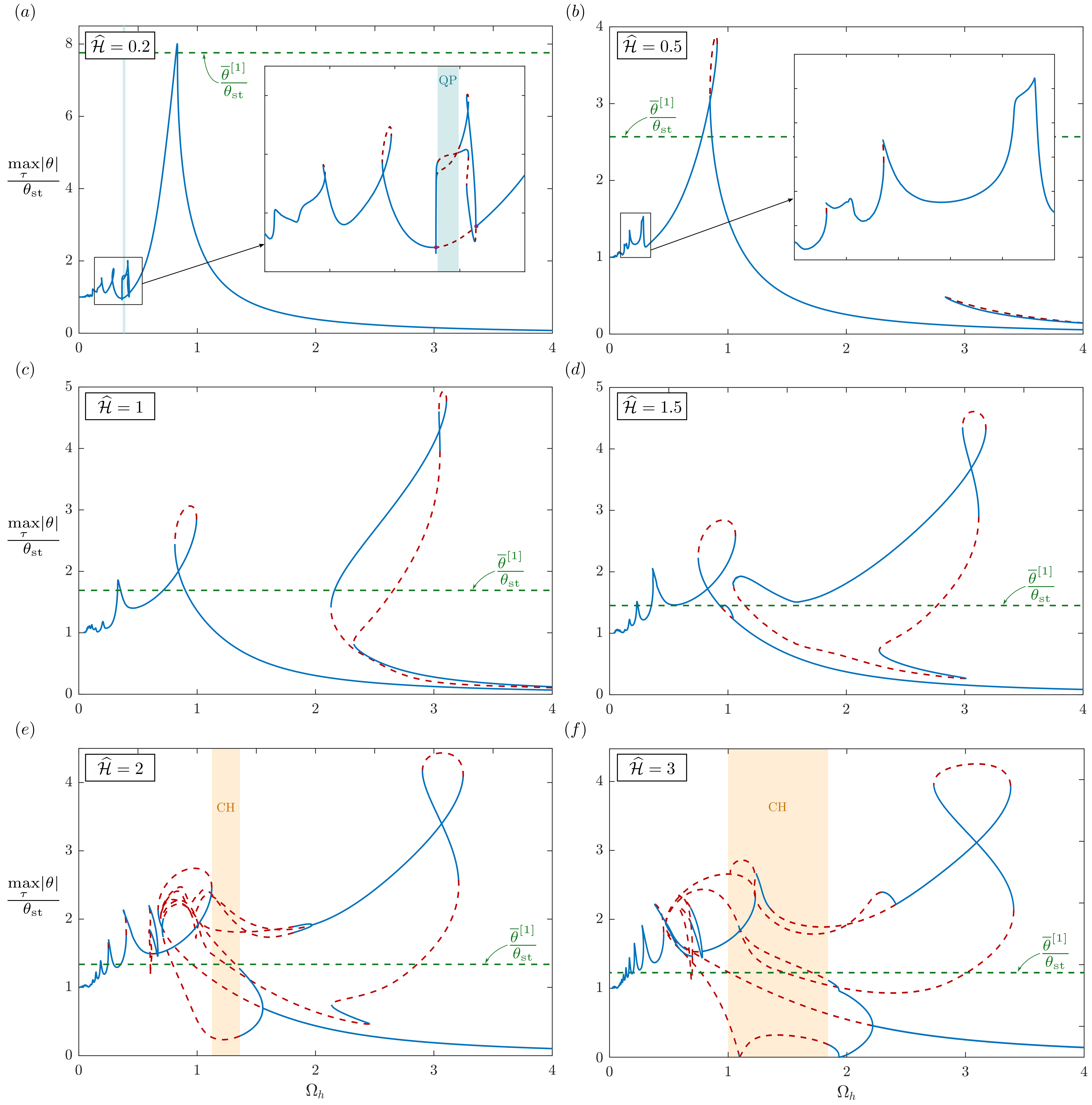}\caption{Normalized maximum absolute value of the difference angle $\theta$ - frequency $\Omega$ response spectrum for the system  with  $\beta_0=80^{\circ}$ and $\mathcal{C}=0.1$ under loading condition (i), eqn. \eqref{loadpatha}, corresponding to harmonic horizontal load  $\mathcal{H}(\tau)= \widehat{\mathcal{H}} \sin (\Omega \tau)$ with amplitude $\widehat{\mathcal{H}}=\{0.2, 0.5, 1, 1.5, 2, 3\}$ at constant vertical  loading $\overline{\mathcal{V}}=-0.5$. The stable (unstable) configurations  are represented as continuous (dashed) blue (red) line on the spectrum. The dashed horizontal green line represents the smallest difference angle $\theta$ for which the parallelogram reduces into a straight line, while the light blue band represents a quasi periodic (QP) response and the light orange band a chaotic (CH) response. }
\label{fig:Response Spectrum}
\end{figure}

The response spectra under  the loading condition (i), eqn. \eqref{loadpatha}, are reported in Fig. \ref{fig:Response Spectrum} with varying the angular frequency $\Omega$ for different horizontal load amplitudes  $\widehat{\mathcal{H}}=\{0.2,0.5,1, 1.5,2,3\}$.
It can be noted that:
\begin{itemize}
    \item for sufficiently small load amplitudes ($\widehat{\mathcal{H}}=0.2$) a classical frequency response curve for the primary resonance is qualitatively detected, which is modified through the emergence of some unstable branches and loops (knots)  when the load amplitude increases ($\widehat{\mathcal{H}}=\{0.5,1,1.5\}$). Remarkably, the self-intersection appearing when a portion of spectrum displays a knot shape is related to  two different stable dynamic steady states with the same amplitude and frequency but shifted in their phase;
    \item super-harmonic 
    resonances are displayed, also with  some unstable branches (see the zoom-in frames for $\widehat{\mathcal{H}}=\{0.2,0.5\}$, but also for higher $\widehat{\mathcal{H}}$) and lead to quasi-periodic motions ($\widehat{\mathcal{H}}=0.2$);
    \item for sufficiently large load amplitudes ($\widehat{\mathcal{H}}=\{0.5,1,1.5,2,3\}$), a  portion of spectrum appears as a loop detached  from the main one defining larger rotation amplitudes and where, due to the loop shape, the jump-phenomena occur at the saddle node equilibria;
    \item a region of chaotic response appears for large load amplitudes ($\widehat{\mathcal{H}}=\{2,3\}$) for a range of frequencies  becoming wider with the increase of $\widehat{\mathcal{H}}$. This can interpreted as the  collision of the coexisting equilibria,  occurring approximately within the frequency range $\Omega\in\left[1,2\right]$. Interestingly,  the system may display the well-known \emph{grazing} phenomenon  \cite{grazing, THOMPSON19825, grazing2, BUDD1995, LYU2020, JIANG2017, YIN2019106}, for which a stable  path can pass through a chaotic region ($\widehat{\mathcal{H}}=3$).
\end{itemize}

\begin{figure}[!ht]
\centering
\includegraphics[width=0.85\textwidth]{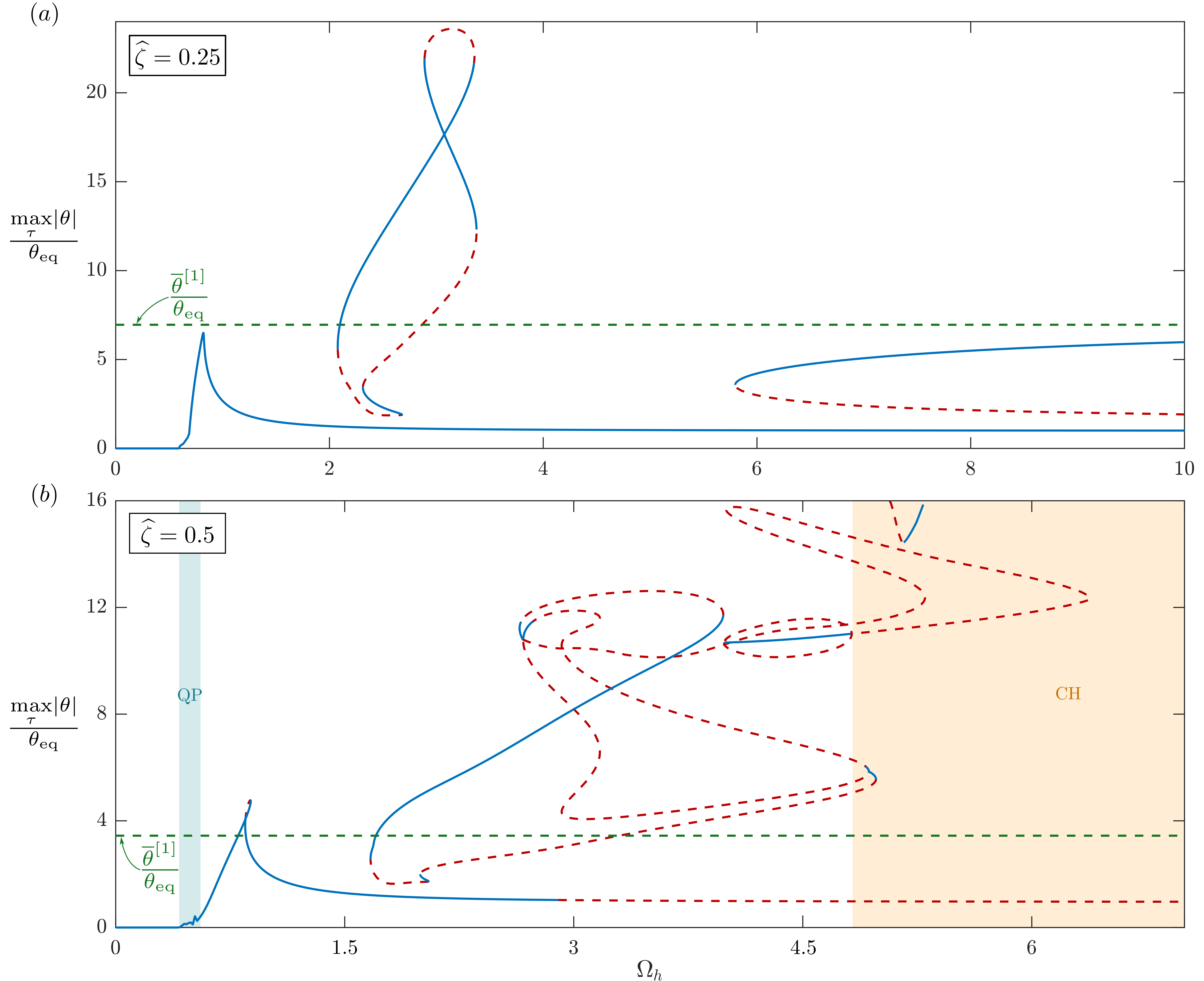}
\caption{ As for Fig. \ref{fig:Response Spectrum}, but under loading condition (ii), eqn. \eqref{loadpathaa}, corresponding to harmonic horizontal ground motion $\zeta(\tau)= \widehat{\zeta} \sin(\Omega \tau)$ with amplitude $(a)$ $\widehat{\zeta}=\{0.25, 0.5\}$
at  constant vertical loading $\overline{\mathcal{V}}=-0.5$. }
\label{fig:Response Spectrum GM}
\end{figure}

The response spectra under the loading condition (ii.), eqn. \eqref{loadpathaa}, are reported in Fig. \ref{fig:Response Spectrum GM} with varying  angular frequency $\Omega$ and for different  ground motion amplitudes $\widehat{\zeta}=\{0.25,0.5\}$.
It can be observed that:
\begin{itemize}
    \item differently from the loading condition (i.), super-harmonic resonances are no longer present, while detached portions of spectrum still appear;
    \item quasi-periodic response may occur at low frequencies as   transition zone between the undeformed state and the steady state ($\widehat{\zeta}=0.5$), as the amplitude of the ground motion increases;
    \item  chaotic response is also detected for high frequencies as $\widehat{\zeta}$ increases.
\end{itemize}

\begin{figure}[!ht]
\centering
\includegraphics[width=\textwidth]{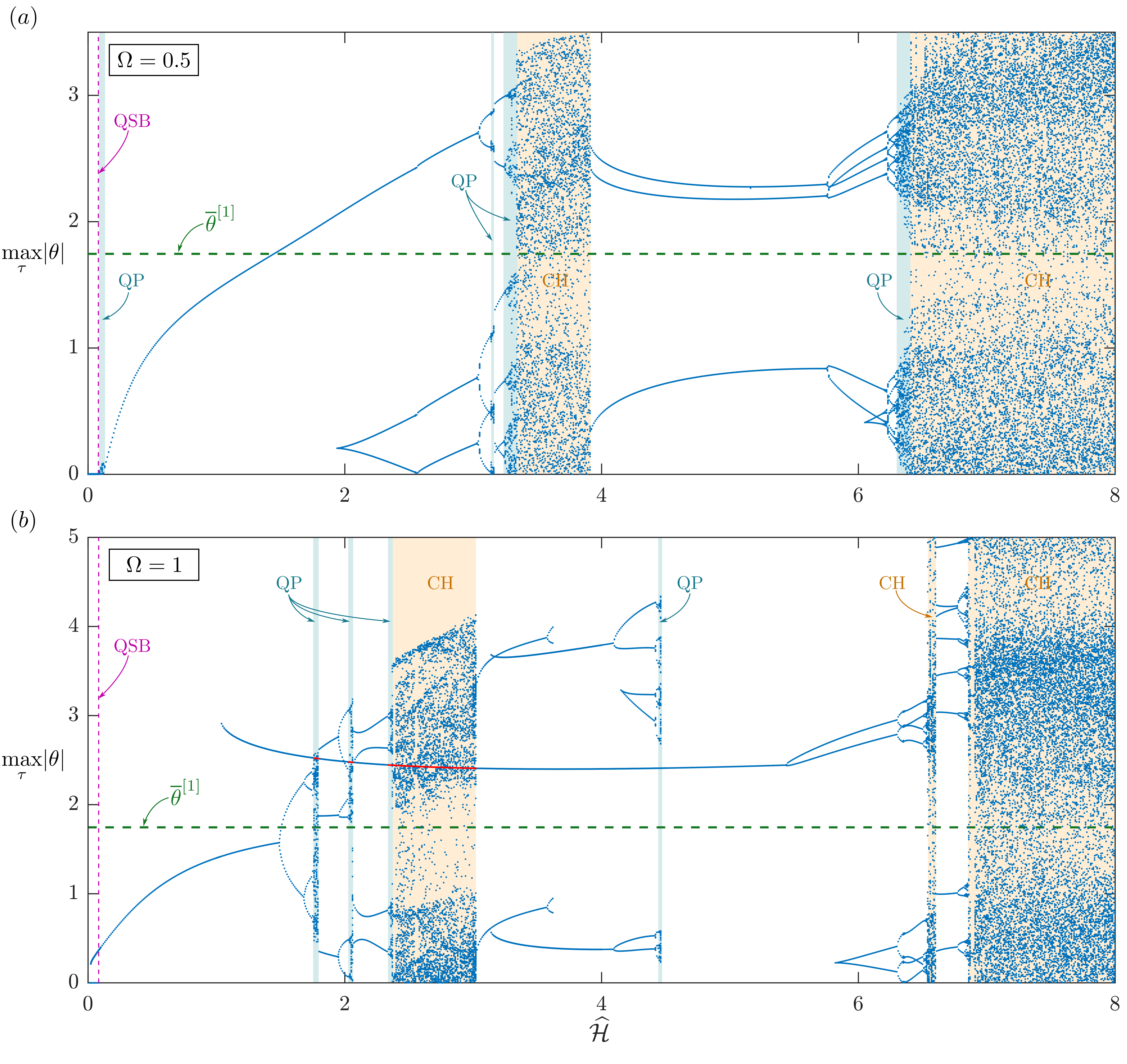}
\caption{Maximum absolute value of the difference angle $\theta$ - horizontal load amplitude $\widehat{\mathcal{H}}$  bifurcation diagram for the system with $\beta_0=80^{\circ}$ and  $\mathcal{C}=0.1$ under loading condition (i), eqn. \eqref{loadpatha}, corresponding to constant vertical  loading $\overline{\mathcal{V}}=-0.5$ and harmonic horizontal loading $\mathcal{H}(\tau)=\widehat{\mathcal{H}} \sin (\Omega \tau)$ and frequency $\Omega=\{0.5, 1\}$.}
\label{fig:Bif_Diagram_1&2}
\end{figure}

\begin{figure}[!ht]
\centering
\includegraphics[width=\textwidth]{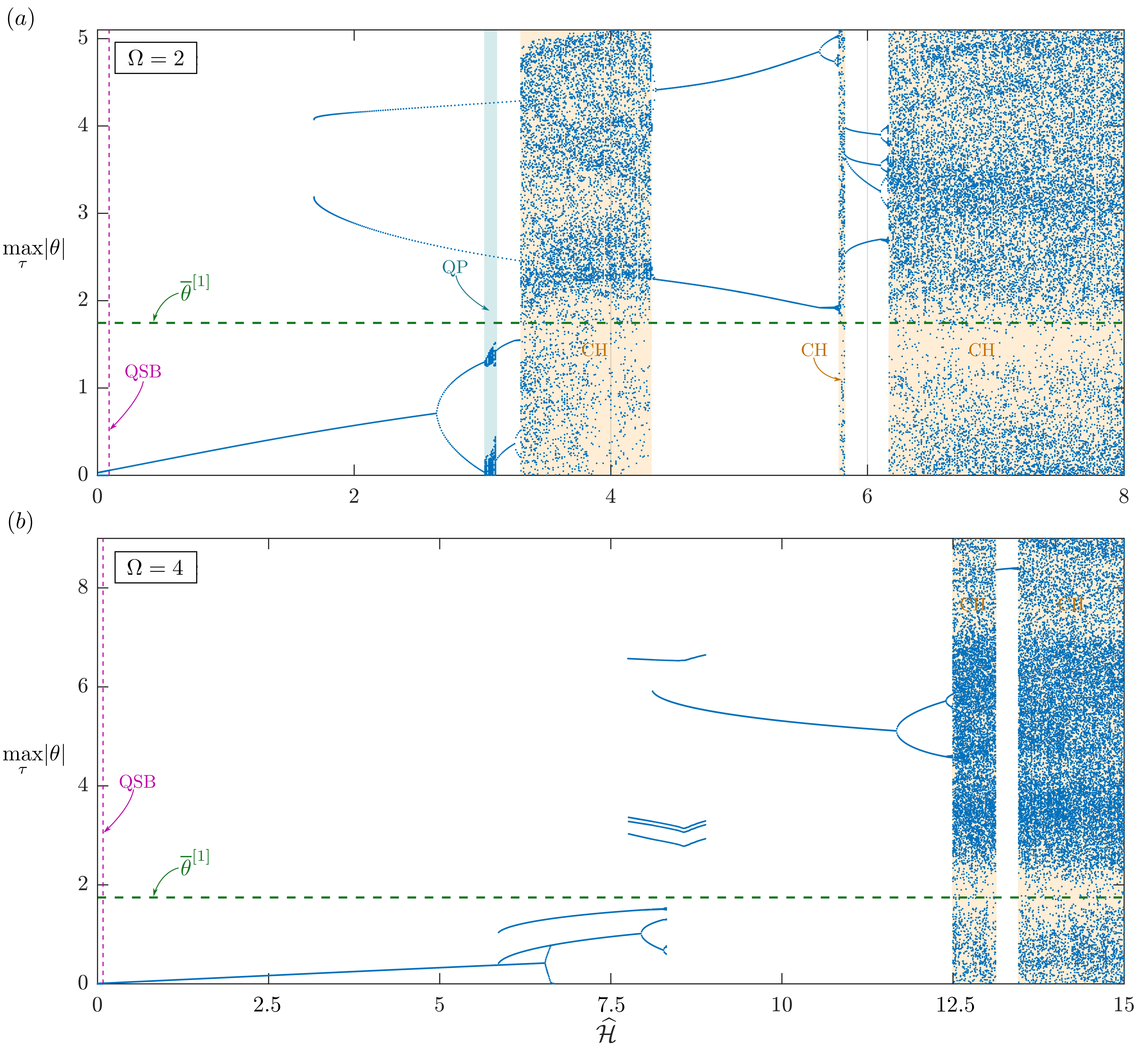}
\caption{As for Fig.  
\ref{fig:Bif_Diagram_1&2}, but for horizontal force frequencies   $\Omega=\{2, 4\}$.}
\label{figBif_diagram_3&4}
\end{figure}

To further assess the dynamic response and the existence of stable attractors, bifurcation diagrams have been obtained under the loading conditions (i) and  (iii)  as maximum  absolute value of rotation $\max_{\tau}|\theta(\tau)|$ with varying the horizontal load amplitude  $\widehat{\mathcal{H}}$ for different values of  frequency $\Omega$.
In particular, the bifurcation diagrams for the loading condition (i) are reported in Fig. \ref{fig:Bif_Diagram_1&2}, for $\Omega=\{0.5,1\}$
and  in Fig. \ref{figBif_diagram_3&4}, for $\Omega=\{2,4\}$. The purple vertical dashed  line represents the quasi-static bifurcation condition (QSB), defined by $\widehat{\mathcal{H}}_{cr}^{QS}
=|\overline{\mathcal{V}}/\tan\beta_0|$.
It can be observed that:
\begin{itemize}
\item the undeformed state ($\theta=0$) becomes unstable  for $\widehat{\mathcal{H}}>\widehat{\mathcal{H}}_{cr}^{QS}$. Moreover,  a supercritical (stable) limit-cycle exists for  $\widehat{\mathcal{H}}<\widehat{\mathcal{H}}_{cr}^{QS}$ with varying the    frequency $\Omega$, providing an anticipation of the bistable response within the quasi-static monostability domain due to dynamics effects (this feature is further analyzed later);
\item the amplitude of the first limit-cycle increases with $\widehat{\mathcal{H}}$ and, for a given $\widehat{\mathcal{H}}$, decreases with $\Omega$, i.e. a flattening of first bifurcated paths manifests itself as the frequency of excitation increases;
\item for sufficiently large $\widehat{\mathcal{H}}$ a very complex dynamic behavior can be detected for which grazing together with a cascade of period doubling bifurcations occur, which constitutes the route for a chaotic response;
\item the chaotic response appears  for two disconnected sets of load amplitudes  $\widehat{\mathcal{H}}$. Indeed,  large-amplitude  stable steady states exist between the two chaotic regions, bifurcating through  period doubling cascade and also grazing \cite{grazing, THOMPSON19825, grazing2, BUDD1995, LYU2020, JIANG2017, YIN2019106} in chaos (see Figs. \ref{fig:Bif_Diagram_1&2}-$a$,$b$);
\item an increase of the frequency $\Omega$ shifts  the region where chaos occurs to larger values of load amplitude $\widehat{\mathcal{H}}$;
 \item the coexistence of multiple stable attractors is analogous to what observed in the response spectra. More specifically, the number of stable attractors increases with the increase of the horizontal load amplitude $\widehat{\mathcal{H}}$. Moreover, the piercing of a stable attractor within a chaotic region (Fig. \ref{fig:Bif_Diagram_1&2}-$b$) corresponds to the presence of stable configuration within the range of frequencies providing chaos  (light orange band)  in the spectra of Figs. \ref{fig:Response Spectrum}-$f$ and  \ref{fig:Response Spectrum GM}-$b$.
\end{itemize}

\begin{figure}[!ht]
\centering
\includegraphics[width=\textwidth]{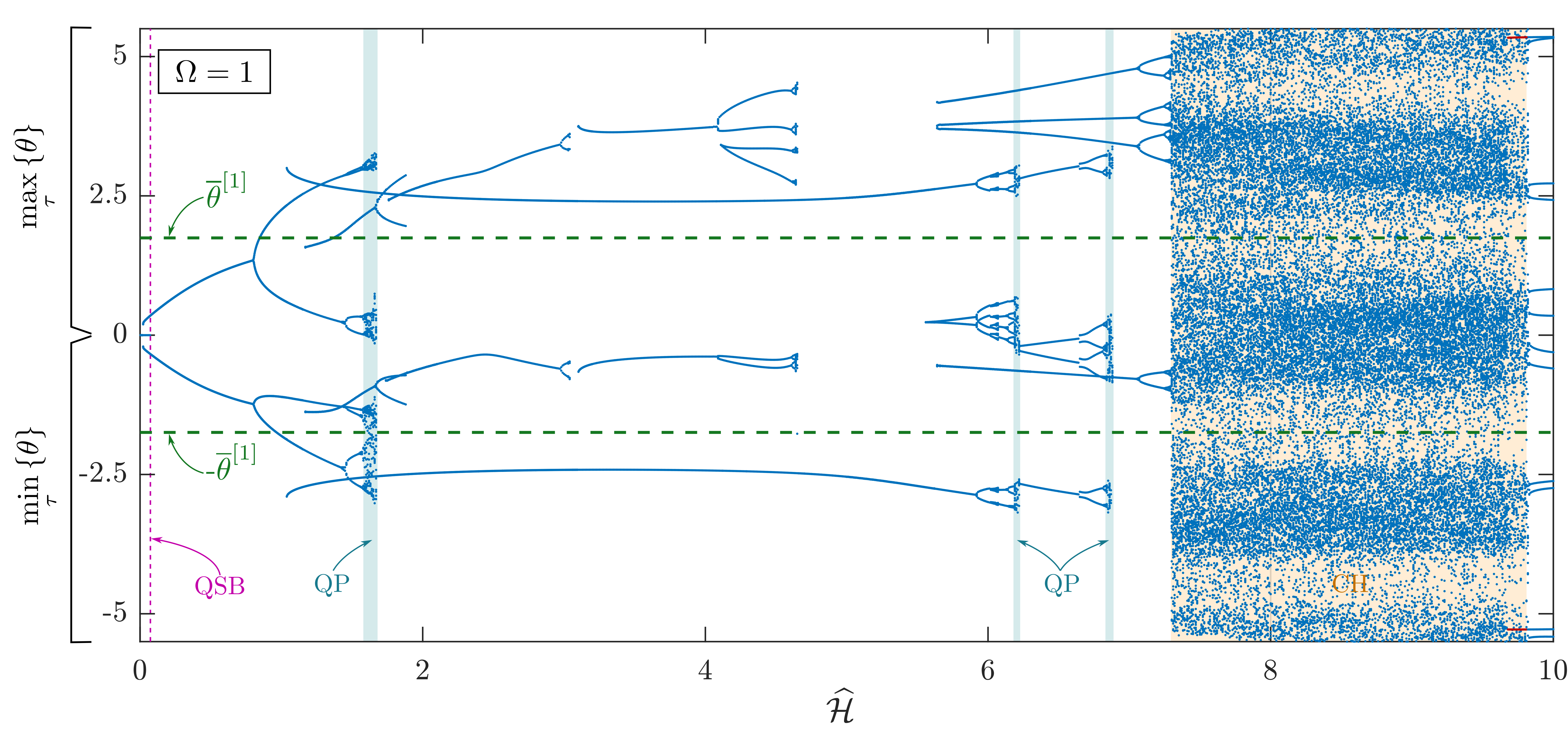}
\caption{As for Fig.  
\ref{fig:Bif_Diagram_1&2}, but under loading condition (iii), eqn. \eqref{loadpathb}, corresponding to harmonic horizontal and vertical forces ($\mathcal{H}(\tau)=\widehat{\mathcal{H}} \sin (\Omega \tau)$, $\mathcal{V}(\tau)=\overline{\mathcal{V}}+\widehat{\mathcal{V}} \cos (\Omega \tau)$) with frequency   $\Omega=1$, vertical force average $\overline{\mathcal{V}}=-0.5$ and amplitude  $\widehat{\mathcal{V}}=0.25$, with varying amplitude of the horizontal  force $\widehat{\mathcal{H}}$. Since the loading is no longer symmetric, the system response loses symmetry and therefore maximum and minimum rotations are reported separately.}
\label{fig:Bif_Diagram_5}
\end{figure}

While  the behaviour is symmetric under loading conditions (i) and (ii), the  dynamic response is no longer symmetric under loading condition (iii), \eqref{loadpathb}, since it is related to an elliptical loading path in the $\mathcal{H}-\mathcal{V}$ plane (Fig. \ref{fig:loading_paths}-$b$),  therefore sensitive to the direction.
Such non-symmetric behavior can be appreciated from the bifurcation diagram reported in terms of maximum  and minimum values of rotation, $\max_{\tau} \{\theta(\tau)\}$ and  $\min_{\tau} \{\theta(\tau)\}$, in Fig. \ref{fig:Bif_Diagram_5} for $\overline{\mathcal{V}}=-0.5$, $\widehat{\mathcal{V}}=0.25$,  and $\Omega=1$. However, apart for the lack of symmetry, the same features and the coexistence of multiple equilibrium paths observed in the symmetric response are also detected here.  

\begin{figure}[!ht]
\centering
\includegraphics[width=\textwidth]{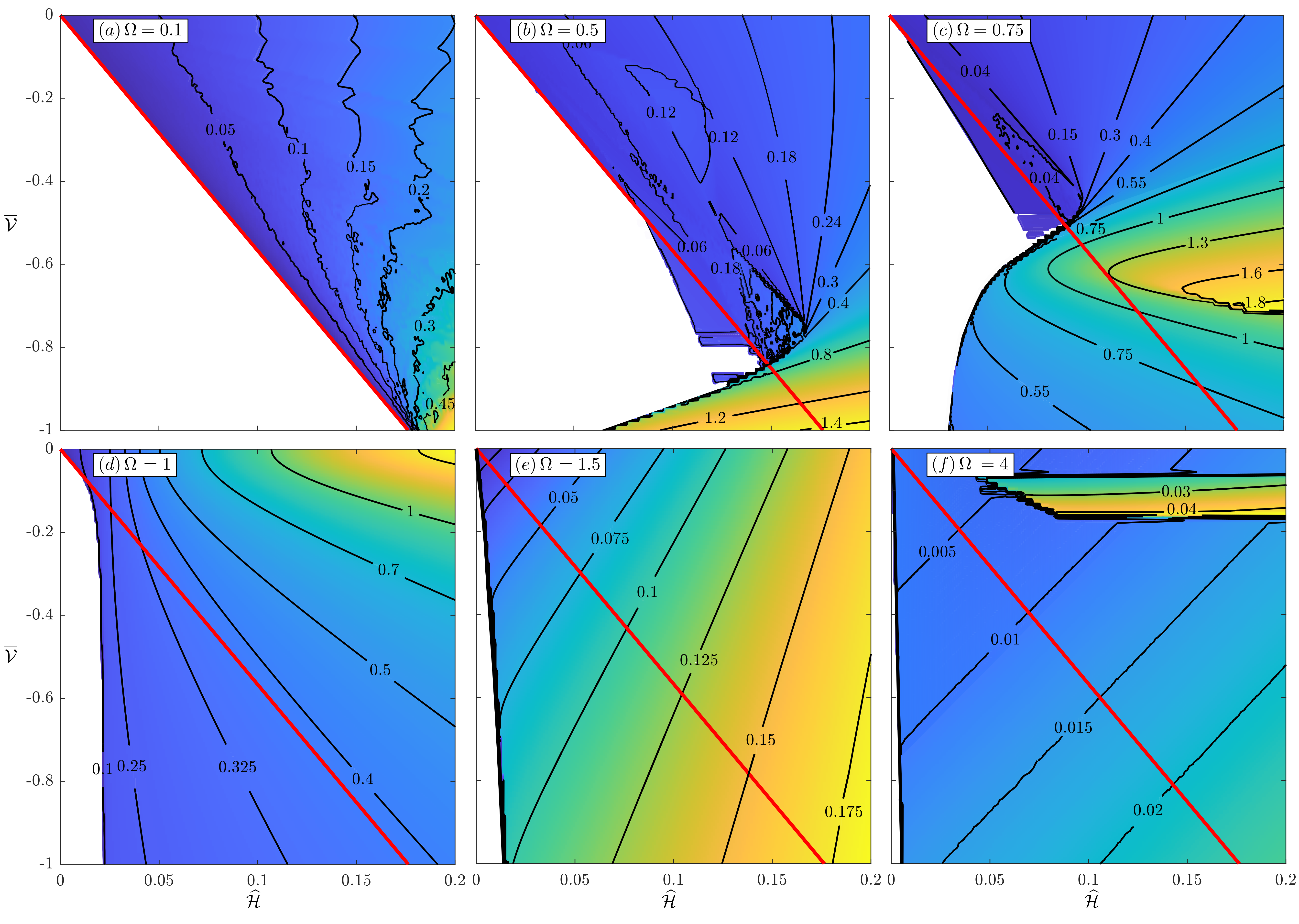}
\caption{Dynamic stability chart for the system with $\beta_0=80^\circ$ and $\mathcal{C}=0.1$ under loading condition (i), eqn. \eqref{loadpatha}, corresponding to harmonic horizontal loading $\mathcal{H}(\tau)=\widehat{\mathcal{H}} \sin (\Omega \tau)$ with varying frequency $\Omega=\{0.1, 0.5, 0.75, 1, 1.5, 4\}$ at constant vertical loading   $\overline{\mathcal{V}}$. White area represents the monostable  region where only the trivial configuration is stable, while the colored region represents the presence of at least one non-trivial stable configuration. The  border of the monostable region under quasi-static conditions is drawn as the red line, showing that dynamics anticipates the bistability. Contour levels display  the largest rotation amplitude $\theta_{\mbox{\scriptsize{max}}}$ of the non-trivial stable solutions.}
\label{fig:Dynamic_instability_1}
\end{figure}

\begin{figure}[!ht]
\centering
\includegraphics[width=\textwidth]{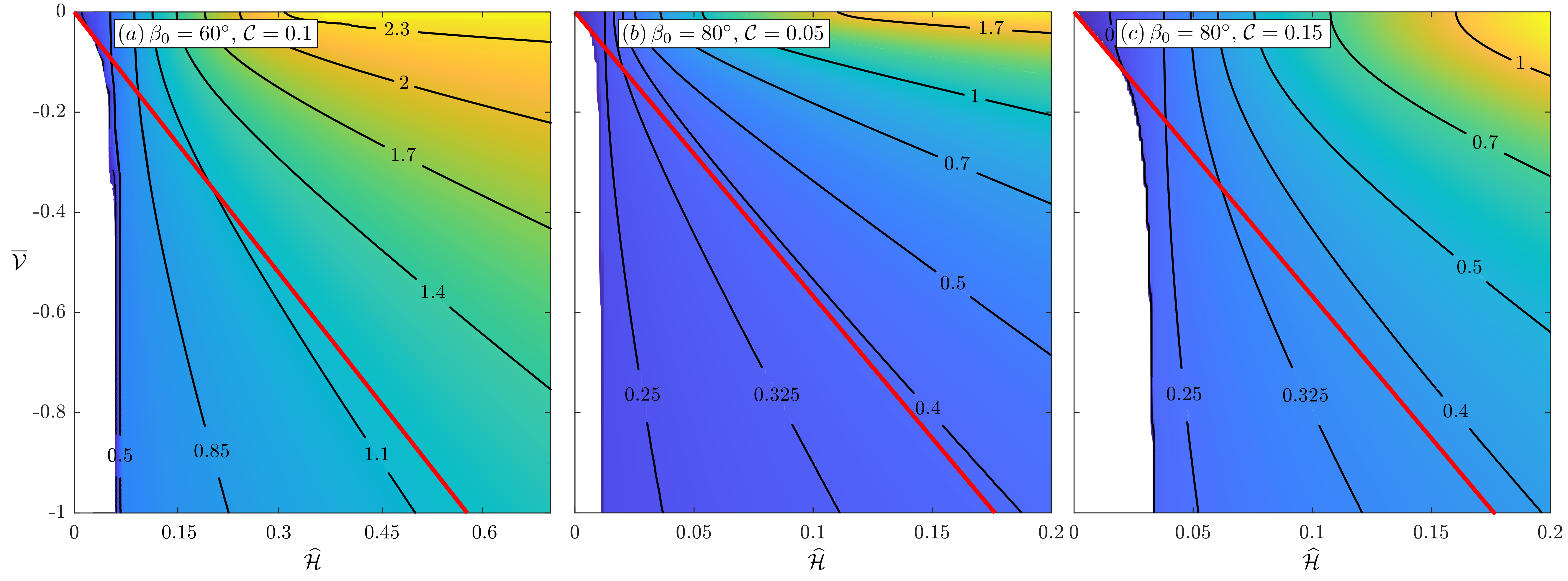}
\caption{As for Fig. \ref{fig:Dynamic_instability_1}, but for a system with $a$) $\beta_0=60^\circ$ and $\mathcal{C}=0.1$, $b$) $\beta_0=80^\circ$  and $\mathcal{C}=0.05$, and $c$) $\beta_0=80^\circ$ and $\mathcal{C}=0.15$, with an horizontal load frequency $\Omega=1$.} 
\label{fig:Dynamic_instability_2}
\end{figure}

A final investigation on   the dynamic behaviour is addressed in terms of the largest rotation amplitude among the non-trivial solutions  under  loading condition (i), eqn. \eqref{loadpatha}. The results are  reported through stability charts in 
Figs. \ref{fig:Dynamic_instability_1} and \ref{fig:Dynamic_instability_2}, realized as contourplots in the 
$\left(\widehat{\mathcal{H}}, \overline{\mathcal{V}}\right)$ plane. White and coloured regions  represent respectively the non-existence and existence of non-trivial stable configurations for the corresponding loading pairs $\left(\widehat{\mathcal{H}}, \overline{\mathcal{V}}\right)$, with the contours associated with the rotation amplitude of the non-trivial stable solution.
Since under quasi-static loading conditions non-trivial equilibrium configurations exist only when $\mathcal{V}\cos\beta_0>-|\mathcal{H}|\sin\beta_0$ (red line), it can be appreciated that  
dynamics reduces the monostability region by increasing the multistability one, therefore realizing an enlargement of the co-existence domain for the  stable trivial and non-trivial  solutions.
In particular, the stability charts reported in Fig. \ref{fig:Dynamic_instability_1} for different frequencies, $\Omega=\{ 0.1,  0.5, 0.75, 1, 1.5, 4\}$, show that the multistability region increases more with the increase of the excitation frequency $\Omega$, while the corresponding amplitudes reach instead highest values for $\Omega\approx 1$. It also noted that a second non-trivial stable motion appears for a narrow region in the  load values $\left(\widehat{\mathcal{H}}, \overline{\mathcal{V}}\right)$ when  $\Omega=4$.

The influence of the initial configuration angle $\beta_0$ and of the damping coefficient  $\mathcal{C}$ can be appreciated from the stability charts for $\beta_0=60^\circ$ and $\mathcal{C}=0.1$ (Fig.\ref{fig:Dynamic_instability_2}-$a$), for $\beta_0=80^\circ$ and  $\mathcal{C}=0.05$ (Fig.\ref{fig:Dynamic_instability_2}-$b$), and for $\beta_0=80^\circ$ and  $\mathcal{C}=0.15$ (Fig.\ref{fig:Dynamic_instability_2}-$c$), all of these for an horizontal load frequency $\Omega=1$.
It is noted  that the initial configuration angle $\beta_0$ has only a slight influence on the shape of the new border (compare Figs. \ref{fig:Dynamic_instability_1}-$d$) and Fig.\ref{fig:Dynamic_instability_2}-$a$)), although  the maximum amplitude decreases with the increase of $\beta_0$. Moreover,  the lower the damping coefficient $\mathcal{C}$, the more the domain of non-trivial solutions extends, as well as the higher the damping coefficient $\mathcal{C}$, the lower the maximum rotation amplitude of the non-trivial solution. 

It is worth to remark that 
the external  load also entails a parametric excitation through a change in the equivalent  stiffness in the  equation of motion \eqref{gov_eqn_constant_loading_lin}. Therefore, the reported stability charts   resembles in a sense  a sort of Strutt-like diagram, although an interplay between the external and the parametric excitation occurs in the present system, differently from the classical Mathieu problem.

\section{Conclusions}

The nonlinear dynamic response has been investigated for a novel metainterface based on two layers of elements  buckling in tension (or equivalently in compression) modeled as a two degree of freedom system, where each degree is subject to a unilateral constraint. 
A system of two nonlinear coupled equations of motion under unilateral constraints have been obtained and, by assuming no bouncing at impact  among all the possible impact scenarios,  reduced to a single nonlinear piecewise-smooth equation of motion, resembling  that governing the rocking oscillations of rigid blocks. Vibrations are then investigated under constant loadings, disclosing divergence condition for small amplitude vibrations, the phase portraits at large amplitude vibrations, and the possibility to realize a bouncing-like motion from a no-bouncing system with  extreme stiffness ratios. Finally, vibration analysis is extended to oscillatory loadings through the presentation of  response spectra, bifurcation diagrams, and stability charts, which show a wide landscape of the metainterface response, encompassing dynamic bifurcations and multistability  anticipation.

The disclosure of such  interesting features within a theoretical framework asks now for an experimental validation and a tuning of dissipation parameters towards an effective implementation of the present metainterface concept in engineering applications ranging from energy harvesting to vibration mitigation.

\section*{Acknowledgements} 
NH gratefully acknowledges the financial support from the European Union’s Horizon 2020 research and innovation programme under the Marie Sklodowska-Curie grant agreement ‘INSPIRE - Innovative ground interface concepts for structure protection’ PITN-GA-2019-813424-INSPIRE.
FDC and FDA gratefully acknowledge financial support from the ERC advanced grant
ERC-ADG-2021-101052956-BEYOND. Support from the Italian Ministry of Education,
University and Research (MIUR) in the frame of the ‘Departments of Excellence’ grant L. 232/2016 is acknowledged.  This work has been developed under the auspices of INDAM-GNFM.

\bibliography{bibliography}
\bibliographystyle{ieeetr}

\appendix

\section{Rocking motion of a rigid rectangular block over an oscillating rigid flat surface}\label{Appen_A}
The rocking motion of a rigid rectangular body has strong similarities with the dynamics of the structural system presented in the main text, including  a previously undisclosed condition providing energy increase at impact. 

The kinetic energy $\mathcal{T}(t)$ of a rigid rectangular block over  an  rigid flat surface, oscillating through a displacement   $U_h(t)=4 R\zeta(t)/3$, is given by
\begin{equation}\label{Kinetic Energy}
	\begin{array}{lll}
		\mathcal{T}(t) =\dfrac{2 M R^2}{3}\left( \dot{\theta}(t)^2  + 4 \dot{\zeta}(t)^2 +2 \dot{\theta}(t) \dot{\zeta}(t) \cos (\alpha -| \theta (t)| )\right) ,
	\end{array}
\end{equation}
where $M$ is the mass of the rigid block, $\alpha$ is the angle between the radius $R$ connecting the center of the mass of the body to its pivoting point and the vertical direction when the  block is at rest, and $\theta(t)$ is the rotation of the rectangular body around the pivoting point.

Since the potential energy $\Pi(t)$ is given as the gravitational potential of the block due to the gravity acceleration  $g$ as 
\begin{equation}\label{Potential Energy}
	\begin{array}{lll}
		\Pi(t) = M g R \left[\cos (\alpha -| \theta (t)| )-\cos \alpha \right],
	\end{array}
\end{equation}
the Euler-Lagrange equation leads to  the celebrated nonlinear equation of rocking motion derived for the first time by Housner \cite{Housner1963} 
\begin{equation}\label{EOM&Rock}
	\begin{array}{lll}
		\ddot{\theta}(t) = -\ddot{\zeta}(t) \cos (\alpha -| \theta (t)| )
		-p^2
		\text{sgn}\left[\theta (t)\right] \sin (\alpha -| \theta (t)| ),
	\end{array}
\end{equation}
where $p^2=3g/(4R)$. 
By substituting $\alpha=\pi/2-\beta_0$, eqn. \eqref{EOM&Rock} can be rewritten as   
\begin{equation}\label{EOM&Rock2}
	\begin{array}{lll}
		\ddot{\theta}(t) = \ddot{\zeta}(t) \sin (| \theta (t)|+\beta_0  )-p^2\text{sgn}\left[\theta (t)\right] \cos (| \theta (t)| +\beta_0),
	\end{array}
\end{equation}
which is similar to equation of motion  \eqref{gov_eqn_theta} for the  structural unit cell  subject to a constant vertical load $\overline{\mathcal{V}}=-p^2$ and $\mathcal{C}=\mathcal{H}(\tau)=\ddot{\nu}(\tau)=0$ and differs only in  the time normalization and in the  linear term in rotation due  to the  presence of the rotational spring.

Next, by considering the   dissipation at impact through the  velocity reduction factor $e=\dot\theta(t^+_*)/\dot\theta(t^-_*)\in(0,1)$,
the jump in the kinetic energy $\mathcal{T}$ at the time of impact $t_*$ follows as
\begin{equation}
   \salto{0.38}{ \mathcal{T}(t_*)} = -\dfrac{2\left(1-e\right)M R^2}{3}\left\{(1+e)  \left[\dot{\theta}(t^-_*)\right]^2 +   2\dot{\theta}(t^-_*)\dot{\zeta}(t_*)\cos \alpha  \right\}.
\end{equation}
This last equation shows that  in the absence of ground motion ($\dot{\zeta}(t_*)=0$) the jump in the kinetic energy is always negative,  $\salto{0.38}{ \mathcal{T}(t_*)}<0$, since $e<1$. However, in the presence of non-null ground  velocity, a kinetic energy decrease is guaranteed only when  the following conditions are both satisfied
\begin{equation}
    \dot{\theta}(t^-_*) \dot{\zeta}(t_*) > 0,\qquad \mbox{and}\qquad 
    |\dot{\zeta}(t_*)|<\dfrac{ 1+e}{2\cos \alpha }\left|\dot{\theta}(t^-_*)\right|.
\end{equation}
Finally, from small vibrations analysis in the absence of ground motion,  when an initial small rotation $\theta_0$ is assumed together with a null initial velocity, the oscillation period for the system in the absence of impact dissipation ($e=1$) is approximated by 
\begin{equation}
    \Te_0(\theta_0)\approx4\sqrt{\dfrac{2 |\theta _0|}{p^2 \sin \alpha }},
\end{equation}
when $\theta_0\ll\alpha$, while it is approximated by
\begin{equation}
    \Te_0(\theta_0)\approx\dfrac{4}{p}\mbox{arccosh}\left[\dfrac{1}{1-\dfrac{|\theta_0|}{\alpha}}\right],
\end{equation}
when  $\theta_0\approx\alpha$, recalling  the overturning initial rotation for the rocking problem $|\theta_0|>\alpha$. It is noted that, by considering  impact dissipation through the velocity reduction factor $e$, the amplitude $\theta_D^{\{n\}}$ and period $\Te_D^{\{n\}}$ of the $n$-th cycle (each cycle encompassing two impacts) and the end  time $t_{\mbox{\scriptsize{fin}}}$ of rocking  are   approximated by the same expressions, eqns. \eqref{approx1}--\eqref{finaltime},  derived for the structural unit cell investigated in the main text.

\end{document}